\documentclass[10pt,
aps, 
prd,
reprint, 
longbibliography, 
letter, 
superscriptaddress,
]{revtex4-1}
\usepackage[utf8]{inputenc}
\usepackage{pstricks}
\usepackage{graphicx}
\usepackage{gensymb}
\usepackage{xcolor}
\usepackage[normalem]{ulem}
\usepackage{tikz}
\usepackage{esvect}
\usepackage{amsmath, bm}
\usepackage{pgfplots}
\usepackage{cancel}
\usepackage{verbatim}
\usepackage{makecell}
\usepackage[symbol]{footmisc}
\usepackage{url}
\usepackage{hyperref}
\hypersetup{colorlinks=true,urlcolor=blue,linkcolor=blue,citecolor=blue}

\begin{document}

\title{Role of Long-Range van der Waals Interaction in the Coefficient of Static Friction}
\author{Ram Narayanan}
\affiliation{Horace Mann School, Bronx, NY, USA}
\email{ram\_narayanan@horacemann.org}
\author{Prachi Parashar}
\affiliation{John A. Logan College, Carterville, IL, USA}
\email{Prachi.Parashar@jalc.edu}
\author{K. V. Shajesh}
\affiliation{School of Physics and Applied Physics, Southern Illinois University--Carbondale, Carbondale, IL, USA}
\email{kvshajesh@gmail.com}
\author{S. Vijayakumar}
\affiliation{Columbia University Medical Center, New York, NY, USA}
\email{sv65@cumc.columbia.edu}
\begin{abstract}
To investigate the role of long-range van der Waals interactions in static friction, we derive an analytic expression for the coefficient of static friction $\mu_s$ between two thin layers of polarizable materials under zero load. For simplicity, we model the surface roughness with sinusoidal corrugations and calculate the interaction energy perturbatively up to the second order in corrugation amplitude. The ratio of corresponding maximum lateral Casimir force to normal Casimir force is defined as the coefficient of static friction, which is found to be independent of the dielectric properties of the materials. It depends on the geometric properties, like interlayer separation, corrugation amplitude, and wavelength of the corrugation. As a proof of concept, our predicted values of $\mu_s$ for the 2D van der Waals materials graphene and hexagonal boron nitride are in reasonable agreement with previously reported values in the literature. This simplistic model could be generalized by incorporating other forces, such as the frequency-dependent contributions of van der Waals interactions and  electrostatic interactions.
\end{abstract}
\maketitle
\newpage
\section{Introduction}
The ubiquity of static friction, the force resisting the tendency of motion of contacting surfaces at rest, in macroscopic and nanoscopic regimes alike has resulted in numerous attempts to investigate its fundamental nature. The earliest works of Guillaume Amontons and Charles-Augustin de Coulomb describe three fundamental tenets of friction forces, two of which apply to static friction. The first tenet states that the maximum static friction force is in direct proportion to the normal force between two surfaces, while the second states that the maximum static friction force is independent of contact area (Ref.~\cite{popovfriction} and references therein).

Arthur Jules Morin coined the term coefficient of friction to describe the proportionality constant not only between the maximum static friction force and the normal force between two surfaces but also to refer to the corresponding quantity when the two surfaces are sliding past each other (the case of kinetic friction) (Ref.~\citep{dowson1978} and references therein).  Morin was the first to extensively tabulate coefficients of friction for material pairs, which was a significant contribution not only to scientists in the emerging field of tribology but also to engineers \citep{dowson1978}. The coefficient of friction would prove to be a valuable concept for understanding friction and applying it to everyday situations \cite{blau2001}. 

Coulomb experimentally observed that static friction was independent of contact area, but he also found that static friction depended on contact time (Ref.~\citep{popovfriction} and references therein). These results were counterintuitive at the time and warranted further investigation. Heinrich Hertz was the first to formally study the effects of external mechanical load on the contact and deformation of a sphere and a plane, with his results indicating that contact area vanished in zero-load conditions (Ref.~\citep{bowden1939} and references therein). Several authors, including Binder, Pedersen, Bowden, and Tabor, building on the work of Hertz, demonstrated that there was a distinction between the apparent contact area, the macroscopic area at which two surfaces appeared to touch, and the real contact area, the smaller, microscopic area at which two surfaces were very close together (Ref.~\citep{bowden1939} and references therein). While the force of static friction did not depend on the apparent contact area, as given by Amontons' second tenet, it did depend on the real contact area \cite{tabor1981}. These results led to further investigation of the factors that influenced the real contact area, and models such as the DMT (Derjaguin-Muller-Toporov) and JKR (Johnson-Kendall-Roberts) theories were proposed to investigate the effects of molecular adhesive interactions on asperity contacts. While these models used different approaches to calculate these interactions, they both showed that adhesive forces play a role in forming contact even in the absence of external load \cite{derjaguin1975, johnson1971}. In addition, Tabor demonstrated the importance of adhesion to the force of friction, as adhesive forces contributed to the real area of contact even in zero-load conditions. He further showed that molecular interactions such as electrostatic and van der Waals forces contribute to surface adhesion between two interacting surfaces \citep{tabor1981, tabor1977}.

Despite these advances in understanding the nature of adhesive forces and their contributions to static friction, the atomic-level properties of materials that affect the coefficient of static friction $\mu_s$ have not been well understood. Current recorded values of $\mu_s$ for different material pairs, such as those found in friction coefficient tables, are empirical values determined using methods ranging from well-known incline experiments to more sophisticated tribometer experiments. These empirical tables of friction coefficients are effective for engineering in the regime of macroscopic contact between conventional materials, allowing engineers to select materials that minimize or maximize $\mu_s$ \citep{blau2001}. However, at the micro-scale level, the dependence of $\mu_s$ on surface features produces varying results due to differences in experimental conditions. Predicting $\mu_s$ for multiple systems of novel or modified materials may thus be tedious, requiring repeated experimentation for each pair of surfaces. For example, determining the coefficient of static friction for heterojunctions, which are of interest in the field of nanoelectronics, currently is only done via experimental or computational methods \citep{mandelli2017}.  A theoretical expression for $\mu_s$ would not only be a valuable tool for predicting the coefficient of static friction for novel material pairs but would also shed light on the atomic-level properties of materials that influence $\mu_s$.

Recently, a few authors have derived expressions for $\mu_s$ and showed that the coefficient of static friction depends on factors such as surface periodicity, surface stiffness, and mean surface separation even in the zero-load case \cite{jahangiri2016, muser2001}. However, these works have not substantially addressed the contributions of fundamental interactions to $\mu_s$. 

In our work, we seek to derive an analytic expression for $\mu_s$ using a first-principles approach that considers the contributions of fundamental interactions. As a first attempt to derive this equation, we would like to use a simplistic case in which only a single fundamental force is considered. One such case is interlayer static friction in clean, dry van der Waals materials \citep{marom2010}. In an undisturbed, static system of two layers within a van der Waals material, van der Waals interactions should predominate over electrostatic and permanent dipole-dipole interactions, as charge transfer has not occurred, and the surfaces are clean. As a result, van der Waals materials provide a unique test case for understanding the contributions of only van der Waals interactions to static friction. 

Although we seek to perform first-principles calculations of the van der Waals forces between two surfaces, calculating these forces using a complete Lifshitz theory approach would be tedious, particularly due to the integrals over the frequencies \citep{lifshitz1956, dzyaloshinskii1961}. In order to simplify our calculations, we consider retarded van der Waals interactions, referred to as Casimir-Polder interactions or long-range van der Waals interactions, in which only the zero-frequency polarizability values contribute to the interaction energy. A notable difference between non-retarded van der Waals interactions and Casimir-Polder  interactions is the dependence of the interaction energy on the separation distance $r$: the incorporation of retardation effects gives the Casimir-Polder interaction energy an  $r^{-7}$ dependence as opposed to the standard $r^{-6}$ dependence of the van der Waals interaction energy \citep{dzyaloshinskii1961}. The effects of relativistic retardation on static friction have not been investigated in depth, as non-retarded van der Waals interactions are considered to predominate over retarded van der Waals interactions at distances of less than 100\,\AA\,\citep{bowden1939}. While our incorporation of retardation effects may prevent us from capturing the full van der Waals interaction energy between the surfaces, it leads to significant simplification in our calculations.

In our model of van der Waals materials, the two layers, modeled as sinusoidally corrugated thin plates, experience both normal and lateral forces due to Casimir-Polder interactions. We consider a zero-load case in which the normal component of the Casimir force is the only normal force acting on the surfaces, and we hypothesize that the lateral component of the Casimir force shall be the predominant source of static friction. Furthermore, we propose that an expression for $\mu_s$ may be derived by taking the ratio of the maximum lateral Casimir force with respect to the normal Casimir force. 

The paper is structured as follows: in Section \ref{framework}, we establish our corrugated-plates model and derive the Casimir interaction energy of the system using a second-order perturbative approximation. In Section \ref{equations}, we use the interaction energy expression to derive the lateral and normal forces acting between the plates, from which we derive $\mu_s$ as the ratio of the maximum lateral force to the normal force. In Section \ref{testing}, we test our derived expression for a few van der Waals materials by calculating a theoretical value of $\mu_s$ and comparing it to previously reported values. A discussion of our findings is presented in Section \ref{discussion}, and summary and conclusions of our results are given in Section \ref{conclusions}.
\section{Interaction Energy Between Corrugated Plates}\label{framework}
The coefficient of static friction between two surfaces, according to the Coulomb model, is the ratio of the maximum static friction force to the normal force.  In our proposal, we attribute the origin of these forces to the long-range van der Waals interactions. To obtain these forces, we first find the Casimir interaction energy between the two surfaces. Then, the variation of the interaction energy with respect to the lateral shift will yield the lateral Casimir force, which we can relate with the static friction force, and the variation of the interaction energy with respect to the separation distance between the two surfaces will give the normal Casimir force, which corresponds to the normal force.

The interaction energy between polarizable materials is described by the Lifshitz theory \cite{lifshitz1956, dzyaloshinskii1961}. For simplicity, we use the long-range van der Waals interaction energy between two polarizable atoms, given in Ref.~\citep{casimir1948}, which is
\begin{equation}
U = -\frac{23\alpha_1\alpha_2\hbar c}{(4\pi)^3r^7}\label{pointpoint},
\end{equation}
where $\alpha_1$ and $\alpha_2$ are the polarizabilities of the two point-sized atoms, and $r$ is the separation distance between the two atoms. A $15\%$ difference between the non-retarded and retarded van der Waals interaction energy at distance less than $51$ nm has been observed \cite{garcion2021}. 

In our configuration, we model the two interacting surfaces as infinitesimally-thin layers of polarizable materials with corrugations as shown in Fig.~\ref{plates}. The plates extend to infinity in the $x$ and $y$ directions. We assume the dielectric functions of these materials to be homogeneous and isotropic. To derive the total interaction energy of this system, we first write the interaction energy between two infinitesimal point-like differential area elements on each surface using Eq.~(\ref{pointpoint}), and then integrate it over the entire surface. The relation between the dielectric function ${\bm \varepsilon}({\bf r};\omega)$and polarizability ${\bm \alpha}$ of a point element at position $\mathbf{r}$ \citep{shajesh2017}, is
\begin{equation}
\frac{{\bm \varepsilon}({\bf r};\omega)}{\varepsilon_0} - {\bf 1} = 4\pi {\bm \alpha}(\omega)\delta({\bf r} - {\bf r}_0)
\end{equation}
where $\varepsilon_0$ is the vacuum permittivity. In the Casimir-Polder limit, it is sufficient to use only the static contribution for the frequency. Furthermore, we will consider the materials to be homogeneous and isotropic. Thus, we will use the equation,
\begin{equation}
4\pi\alpha = \left(\frac{\varepsilon}{\varepsilon_0} - 1 \right)dxdydz.\label{dielectricFunction}
\end{equation}Since the materials have negligible thickness, we shall define the two-dimensional polarizability-densities $\sigma_i$, where $i=1,2$, which are related to their respective polarizabilities by 
\begin{eqnarray}
\alpha_i &=& \sigma_i dxdy,
\end{eqnarray}
such that
\begin{eqnarray}
4\pi\sigma = \left(\frac{\varepsilon}{\varepsilon_0} - 1\right)dz\label{sigmadielectric}.
\end{eqnarray}
\\
\subsection{Surface Corrugations Model of van der Waals Interactions}

\begin{figure}[t]
    \vspace{2em}
    \includegraphics[width=8cm]{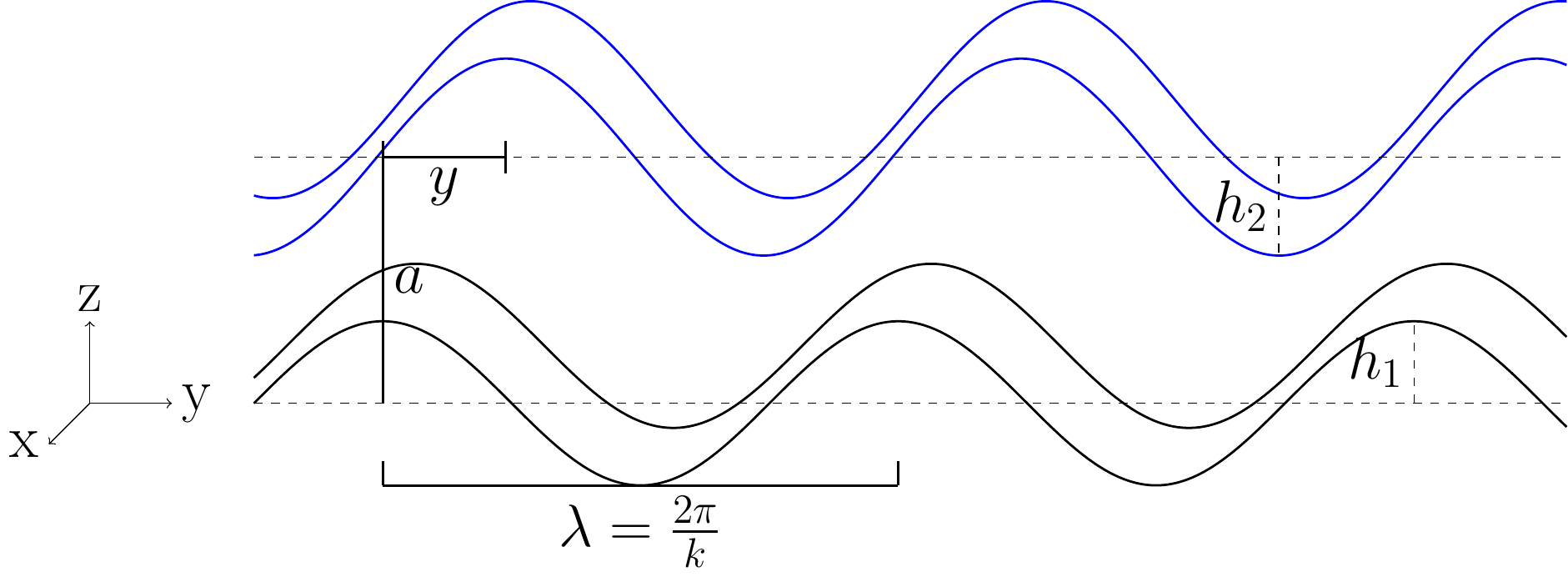}
    \caption{Sinusoidally corrugated plates of wavelength $\lambda$, separated by distance $a$, with corrugation heights $h_1$ and $h_2$ for the bottom and top plates respectively, and the top plate shifted by distance $y$. The corrugations are along the $y$-axis, and the plates extend along the $x$-axis as well.}
    \label{plates}
\end{figure}
To model the surface roughness, we use a sinusoidal corrugation for simplicity. The position of each surface in the $z$-direction is defined as
\begin{subequations}
\begin{flalign}
z_1&= h_1\cos(k y^\prime),\\
z_2&= a + h_2\cos(k (y^{\prime\prime}-y)),
\end{flalign}%
\label{corrugations}%
\end{subequations}%
where we choose the same wavenumber $k=2\pi/\lambda$ for each surface. The amplitudes of the corrugations are $h_1$ and $h_2$, and $y$ represents the lateral shift between the two surfaces, such that the vertical distance between elements in the two layers is
\begin{equation}
r_z = a - h_1\cos(ky') + h_2\cos(k(y'' - y)).\label{corrugationsnew}
\end{equation}
Our corrugated plates model is shown in Fig.~\ref{plates}.
\subsection{Interaction Energy}
The interaction energy $dU$ between differential point elements on the two surfaces is
\begin{eqnarray}
dU = -\frac{23\sigma_1\sigma_2\hbar c}{(4\pi)^3}\frac{dx'dx''dy'dy''}{r^7}.
\end{eqnarray}
where we define, 
\begin{equation}
r^2 = (x^{\prime\prime} - x^{\prime})^2 + (y^{\prime\prime} - y^\prime)^2 + r_z^2.
\end{equation}
The total interaction energy between the two plates may be obtained as
\begin{eqnarray}
U = -\frac{23\sigma_1\sigma_2\hbar c}{(4\pi)^3}\int\frac{dx'dx''dy'dy''}{r^7},
\end{eqnarray}
where the integration is over the entire area of both plates. We shall evaluate our integral using the perturbative expansion as done in Ref.~\citep{noncontactgears1}.
The resulting interaction energy may be grouped into six terms, clubbed by their powers in $h_1$ and $h_2$,
\begin{flalign}
U = U^{(0,0)} + U^{(1,0)} + U^{(0,1)} + U^{(2,0)} + U^{(0,2)} + U^{(1,1)},
\end{flalign}
where the first and second terms in the superscript refer to the order of the perturbative parameters $h_1/a$ and $h_2/a$ of the two plates, respectively. The leading order contributions are
\begin{subequations}
\label{E}
\begin{eqnarray}
U^{(0, 0)} &=& -\frac{23\sigma_1\sigma_2\hbar c}{(4\pi)^3}\int d^4(x, y)\left[\frac{1}{r^7_0}\right],\label{U00-def}
\\U^{(1,0)} &=& -\frac{23\sigma_1\sigma_2\hbar c}{(4\pi)^3}\int d^4(x, y)\left[\frac{7h_1a\cos(ky')}{r^9_0}\right],\label{U10-def}
\\U^{(0,1)} &=& -\frac{23\sigma_1\sigma_2\hbar c}{(4\pi)^3}\int d^4(x, y)\left[-\frac{7h_2a\cos(k(y'' - y))}{r^9_0}\right],\label{U01-def}\nonumber
\\&&\hspace{2cm}
\end{eqnarray}
and the next-to-leading contributions are
\begin{widetext}
\begin{eqnarray}
U^{(2,0)} &=& -\frac{23\sigma_1\sigma_2\hbar c}{(4\pi)^3}\int d^4(x, y)\left[-\frac{7h_1^2\cos^2(ky')}{2r^9_0} + \frac{63h_1^2a^2\cos^2(ky')}{2r^{11}_0}\right],\label{U20-def}
\\U^{(0,2)} &=& -\frac{23\sigma_1\sigma_2\hbar c}{(4\pi)^3}\int d^4(x, y)\left[-\frac{7h_2^2\cos^2(k(y'' - y))}{2r^9_0} + \frac{63h_2^2a^2\cos^2(k(y'' - y))}{2r^{11}_0}\right],\label{U02-def}
\\U^{(1,1)} &=& -\frac{23\sigma_1\sigma_2\hbar c}{(4\pi)^3}\int d^4(x, y)\left[\frac{7h_1h_2\cos(ky')\cos(k(y'' - y))}{r^9_0} -\frac{63h_1h_2a^2\cos(ky')\cos(k(y'' - y))}{r^{11}_0}\right],\label{U11-def}
\end{eqnarray}
\end{widetext}
\end{subequations}
where we have defined 
\begin{equation}
r^2_0 = (x'' - x')^2 + (y'' - y')^2 + a^2
\end{equation}
and
\begin{equation}
d^4(x, y) = dx'dx''dy'dy''
\end{equation}
for brevity. 

One could alternatively set up this problem with two thin flat plates of sinusoidally varying surface polarizabilities. In that setup, $U^{(1, 0)}$, $U^{(0, 1)}$, $U^{(2, 0)}$, and $U^{(0, 2)}$ could be thought of as arising from the interaction energies between two flat plates, one in which the polarizability is constant across the surface, and another where the surface polarizability varies sinusoidally in the $y$-direction. $U^{(1, 1)}$ may similarly be considered as arising from the interaction energies between two plates, both of which have their polarizabilities varying sinusoidally in the $y$-direction.

The general integral of  $d^4(x, y)/r^n_0$ converges for $n > 1$ as
\begin{equation}
\int \frac{d^4(x, y)}{(x'' - x')^2 + (y'' - y')^2 + a^2)^n} = \frac{L_xL_y\pi}{(n - 1)a^{2(n - 1)}},\label{1Rn}
\end{equation}
where $L_x$ and $L_y$ are the extracted infinite lengths from the integrals over $dx'$ and $dy',$ respectively. 
Using Eq.~(\ref{1Rn}), we can solve the different pieces of the total interaction energies in Eq.~(\ref{E}).

Solving for $U^{(0, 0)}$ in Eq.~(\ref{U00-def}), we have
\begin{eqnarray}
\frac{U^{(0,0)}}{L_xL_y} &=& -\frac{23\sigma_1\sigma_2\hbar c}{(4\pi)^3a^5}\frac{2\pi}{5},\label{U00-solved}
\end{eqnarray}
where we use the substitution $n = 7/2$ in Eq.~(\ref{1Rn}). Note that this is the interaction energy of two flat plates with the same values of $\sigma_1, \sigma_2$ and $a$. Further results for pieces of the interaction energy shall be written as this expression multiplied by a dimensionless factor. The interaction energy associated with the first-order term in $h_1$ in Eq.~(\ref{U10-def}),
\begin{equation}
U^{(1, 0)} = 0.
\end{equation}
This result is due to the isolated $\cos(ky')$ integral, which may be broken into discrete pieces of one complete period,
\begin{equation}
\int_{-\infty}^{\infty}\cos(kx)dx = \sum_{n=-\infty}^{\infty}\frac{k}{2\pi}\int_{0}^{\frac{2\pi}{k}}\cos(kx)dx = 0.\label{cosInt}
\end{equation}
Likewise,
\begin{equation}
U^{(0, 1)} = 0.
\end{equation}
The four integrals of the interaction energies associated with the second-order terms in $h_1$ and $h_2$ may each be broken up into isolated square-of-cosine integrals and $1/r^n_0$ integrals. The results for $U^{(2, 0)}$ and $U^{(0, 2)}$ are thus
\begin{eqnarray}
\frac{U^{(2,0)}}{L_xL_y} = -\frac{23\sigma_1\sigma_2\hbar c}{(4\pi)^3a^5}\,\frac{15}{2}\pi\frac{h_1^2}{a^2},\\
\frac{U^{(0,2)}}{L_xL_y} = -\frac{23\sigma_1\sigma_2\hbar c}{(4\pi)^3a^5}\,\frac{15}{2}\pi\frac{h_2^2}{a^2}.\label{U02-solved}
\end{eqnarray}
The two terms of $U^{(1, 1)}$ may be solved with the substitution $y''' = y' - y''$, isolating the $y''$ cosine terms and giving only $\cos(ky''')$ in the numerator of the $dy'''$ integral due to cancellation of the odd sine integrals. The resulting integrands of the $dy'''$ integral may be solved using the following integration formulae in succession \citep{gradshteyn2007},
\begin{eqnarray}
\,&&\int_{0}^{\infty}\frac{x^{2m}\cos(kx)dx}{(x^2 + a^2)^{n + \frac{1}{2}}} \nonumber\\
&&\hspace{1cm}= \frac{(-1)^m\sqrt{\pi}}{2^na^n\Gamma{(n + \frac{1}{2})}} \cdot \frac{d^{2m}}{dk^{2m}}\left[k^nK_n(ka)\right],\label{U11CosInt}\\
\,&&\int_{0}^{\infty}\frac{K_\nu(\alpha\sqrt{x^2 + z^2})x^{2\mu+1}dx}{\sqrt{(x^2 + z^2)^{\nu}}} \nonumber \\
&&\hspace{1cm}= \frac{2^\mu\Gamma(\mu + 1)K_{\nu - \mu - 1}(\alpha z)}{\alpha^{\mu + 1}z^{\nu-\mu-1}},\label{bessInt}
\end{eqnarray}
where $K_n$ is the modified Bessel function of the second kind. This gives
\begin{eqnarray}
\frac{U^{(1, 1)}}{L_xL_y} &=& \frac{23\sigma_1\sigma_2\hbar c}{(4\pi)^3a^5}\frac{e^{-ka}\pi\cos(ky)}{15}\frac{h_1}{a}\frac{h_2}{a}\\
&\times& \left[(ka)^4 + 9(ka)^3 + 39(ka)^2 + 90(ka) + 90\right].\nonumber
\end{eqnarray}
\section{Coefficient of Static Friction}\label{equations}
Using the expressions from the previous sections, the total Casimir interaction energy per unit area between two sinusoidally corrugated plates up to the second-order in the perturbative parameters $h_i/a$,
\begin{eqnarray}
\frac{U}{A}
&=& -\frac{23\sigma_1\sigma_2\hbar c\pi}{(4\pi)^3a^5}\frac{2}{5}\hspace{5.5cm}\nonumber\\
&&\hspace{-0.5cm}\times\left[1 + \frac{15}{2}\frac{h^2_1 + h^2_2}{a^2} - \frac{e^{-ka}\cos(ky)}{6}\frac{h_1}{a}\frac{h_2}{a}P(ka)\right]  ,\label{finalU}
\end{eqnarray}
where $A = L_xL_y$ is the area and we have defined the dimensionless polynomial quantity
\begin{equation}
P(ka) = (ka)^4 + 9(ka)^3 + 39(ka)^2 + 90(ka) + 90
\end{equation}
for brevity.

The lateral force between the plates is the negative partial derivative of the interaction energy with respect to the lateral displacement $y$. Similarly, the normal force between the two plates is the negative partial derivative of the same interaction energy with respect to the vertical displacement $a$. The lateral force per unit area is,
\begin{eqnarray}
\frac{F_L}{A} &=&  \frac{23\sigma_1\sigma_2\hbar c\pi}{(4\pi)^3a^6}\frac{e^{-ka}\sin(ky)}{15}\frac{h_1}{a}\frac{h_2}{a}(ka)P(ka).\label{lateral1}
\end{eqnarray}
The maximum lateral force is given by the maximum value of this sine function, which is
\begin{equation}
\frac{F_{L,\text{max}}}{A} =   \frac{23\sigma_1\sigma_2\hbar c\pi}{(4\pi)^3a^6}\frac{e^{-ka}}{15}\frac{h_1}{a}\frac{h_2}{a}(ka)P(ka).\label{maxLateral}
\end{equation}
Similarly, the normal force between the two plates is the negative partial derivative of the same interaction energy with respect to the vertical displacement $a$. The normal force per unit area is thus,
\begin{equation}
\frac{F_N}{A} = -\frac{46\sigma_1\sigma_2\pi\hbar c}{(4\pi)^3a^6}\left[1 + \mathcal{O}\left(\frac{h^2}{a^2}\right)\right].\label{normal1}
\end{equation}
The exact expression for the normal force contributions of second order in $h/a$, represented by the second term in the square brackets of Eq.~(\ref{normal1}), may be found in Eq.~(\ref{normalfull}). 

We define the coefficient of static friction $\mu_s$ as $\mu_s = \left|F_{L,\text{max}}/F_N\right|$. The concept of using the maximum value of the lateral force as the maximum value of the static friction force has been proposed previously in the regime of atomic-scale friction \citep{zhong1990}. In addition, our model shows that the lateral force is a sine function with respect to the lateral displacement $y$. Thus, when a constant force is applied to the plates that is less than or equal to the maximum lateral force, there will be some value of $y$ for which the lateral force is equal in magnitude and opposite in direction to the applied force. Therefore, the maximum lateral force is the threshold force that must be applied for sliding to occur, which is used to define the coefficient of static friction.
The magnitude of the coefficient of static friction up to the second order in $h/a$ is obtained by keeping the first term of Eq.~(\ref{normal1}),
\begin{eqnarray}
\mu_s &=& \frac{h_1}{a}\frac{h_2}{a}\frac{e^{-ka}}{30}(ka)P(ka).
\label{equationmus}
\end{eqnarray}
Note that the area dependence, along with dielectric dependence of $\sigma_1, \sigma_2$, vanishes in the equation for $\mu_s$ regardless of whether the normal force is approximated to the first term of Eq.~(\ref{normal1}) or not. On the other hand, while Eq.~(\ref{equationmus}) appears to show that $\mu_s$ is independent of $y$, this independence only holds when Eq.~(\ref{normal1}) is approximated to the dominant contribution.
\begin{figure}
    \centering
    \includegraphics[width=8.6cm]{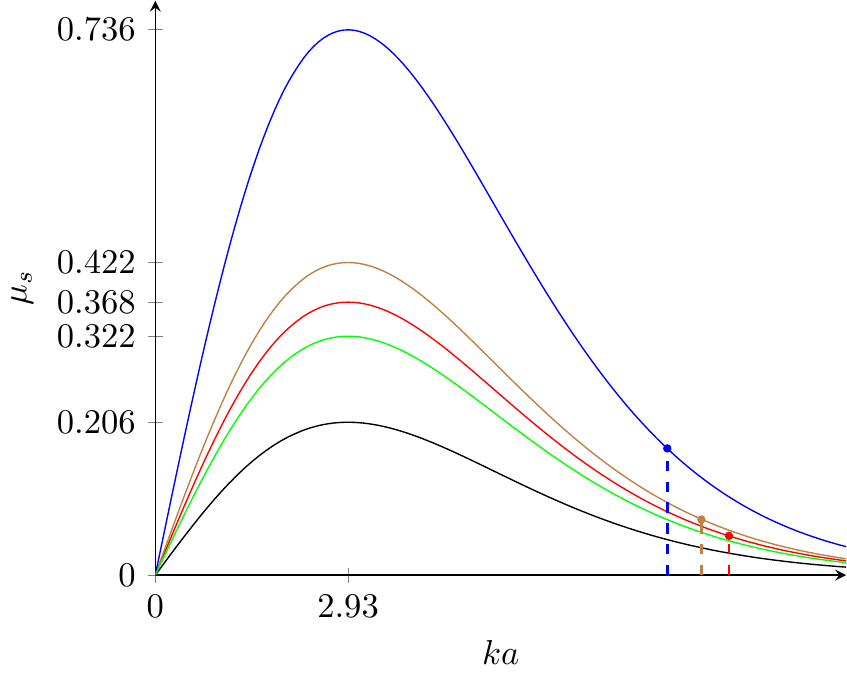}
    \caption{From bottom to top (color online): plots of $\mu_s$ with respect to $ka$ for $h_1/a = h_2/a = 1/5$ (black), $h_1/a = h_2/a = 1/4$ (green), $h_1/a = h_2/a = 0.267$ (red), $h_1/a = 0.277$ and $h_2/a = 0.295$ (brown), and $h_1/a = h_2/a = 0.378$ (blue). The maximum of each curve is marked on the vertical axis. The values of $h_1/a$ and $h_2/a$ for the black and green curves are arbitrarily chosen for comparison. The red, brown, and blue curves correspond to the calculated values of $h_1/a$ and $h_2/a$ for graphene, h-BN, and black phosphorene, respectively, which are the materials discussed in Section~\ref{testing}. For each material, the vertical dotted line from the $x$-axis to a point on the curve represents $\mu_s$ based on the calculated value of $ka$ for these materials.  Note that $\mu_s$ is maximized for $ka \approx 2.93$ independently of $h_1$ and $h_2$.}
    \label{mus1}
\end{figure}
The expression for $\mu_s$ given by Eq.~(\ref{equationmus}) depends on three dimensionless parameters: $h_1/a$, $h_2/a$ and $ka$. A plot of this $\mu_s$ expression with respect to $ka$ is shown with corresponding values of $h_1/a$ and $h_2/a$ for the three materials discussed in Section~\ref{testing}, as well as two arbitrarily chosen values $h_1 = h_2 = a/5$ and $h_1 = h_2 = a/4$, in Fig.~\ref{mus1}.

These plots show that $\mu_s$ is maximized for $ka \approx 2.93$ regardless of $h_1$ and $h_2$. In addition, $\mu_s$ decays as $ka \to 0$ or $ka \to \infty$. 

With respect to each of $h_1/a$ and $h_2/a$, $\mu_s$ exhibits linear scaling and clearly approaches 0 as either $h_1 \to 0$ or $h_2 \to 0$. While $\mu_s$ is always maximized for $ka = 2.93$, the value of this maximum is also determined by $h_1/a$ and $h_2/a$ as demonstrated by the different curves of Fig.~\ref{mus1}. Assuming the conservative bound $h_1 \leq a/2 $ and $h_2 \leq a/2$, which ensures that the corrugated plates do not intersect for any value of the lateral shift $y$, Eq.~(\ref{equationmus}) implies an upper bound of $1.29$ for the coefficient of static friction.

These results seem to suggest that the corrugation wavelength $\lambda$ can be considered as an approximation parameter, as $\mu_s$ may also be written as a function of the dimensionless parameters $h_1/\lambda$, $h_2/\lambda$, and $a/\lambda$.
\section{$\mu_s$ for van der Waals Materials}\label{testing}
Using the analytic expressions derived in the previous section, we calculate the Casimir interaction energy, lateral force, normal force, and coefficient of static friction between layers of graphene, boron nitride, and black phosphorene. Our use of the Casimir-Polder limit results in the interaction energy per unit area defined in Eq.~(\ref{finalU}) scaling with $a^{-5}$ as opposed to the $a^{-4}$ scaling characteristic of the van der Waals interaction energy. Considering that we have used the perturbative approximation in $h_i/a$ and the normal force approximation to the dominant term in addition to the Casimir-Polder limit, we suspect that our results for the interaction energy may have an error margin of $15$ to $20$ percent. 
\subsection{Graphene}
\begin{figure}[hbtp]
    \centering
    \includegraphics[width=8.6cm]{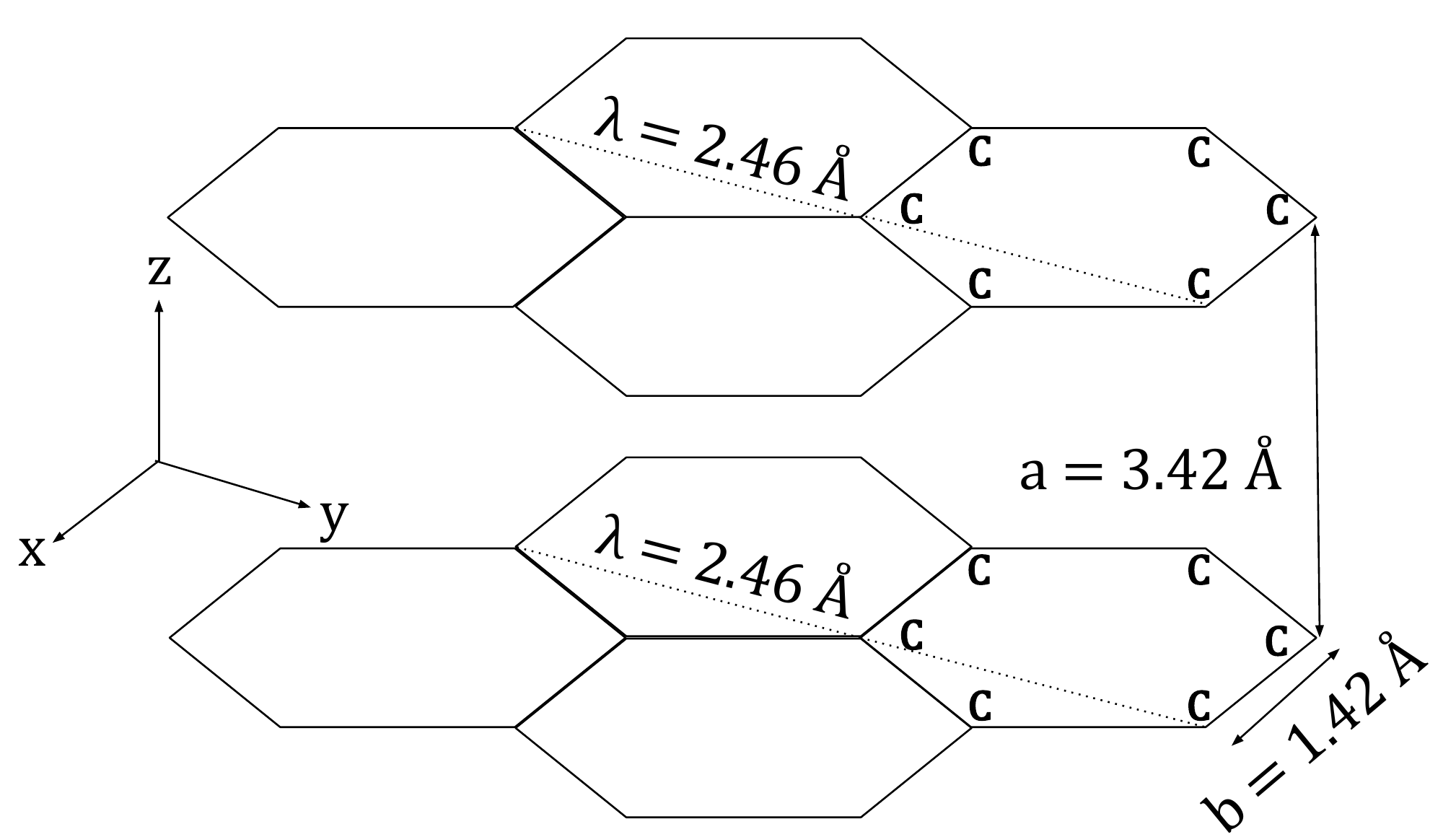}
    \caption{The structure of $AA$-stacked graphene, with the corrugation path being along the short diagonal.}
    \label{grapheneMultilayer}
\end{figure}

\begin{figure}[hbtp]
    \centering
    \includegraphics[width=8.6cm]{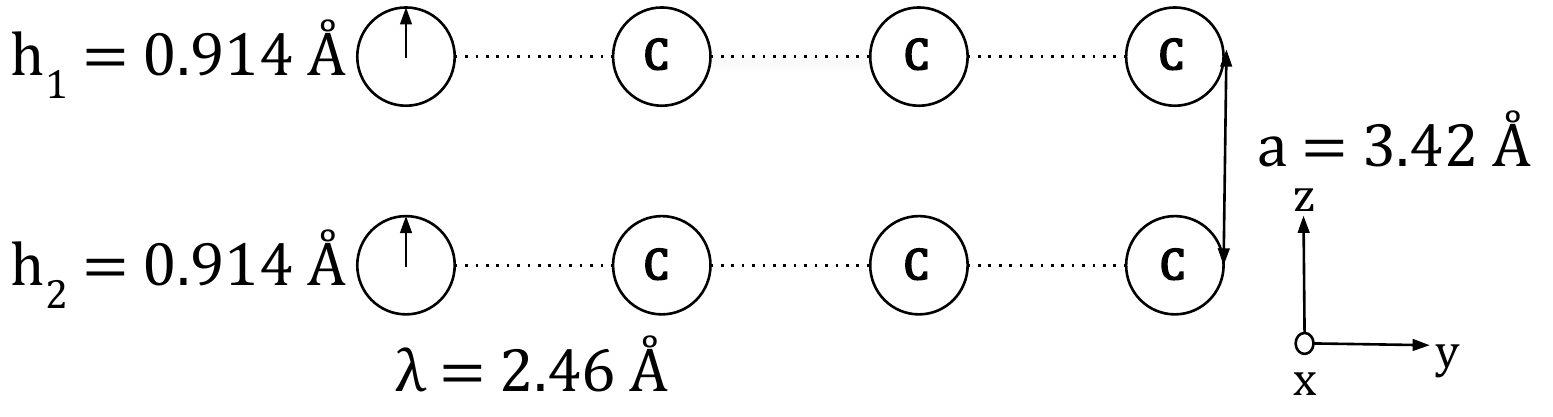}
    \caption{A cross sectional slice showing the short diagonal periodicity, which is treated as a sinusoidal corrugation in the graphene interaction. The carbon atoms' radii are used as the corrugation heights.}
    \label{grapheneInteraction}
\end{figure}

We begin with the graphene-graphene interaction. Figures~\ref{grapheneMultilayer} and~\ref{grapheneInteraction} illustrate the values of $a, \lambda,$ and $h$ used in our calculations for two graphene layers.

For our calculations, we use a polarizability per unit cell of $0.949\,{\text{\AA}\,^3}$ \citep{kumar2016} and a bond length of $1.42\,\text{\AA}\,$\citep{chung2002}, which gives polarizability per unit area $\sigma = 0.181$ \AA. 

Depending on the methodology used, different values have been obtained in the literature for the interlayer distance, $a$, between graphene layers. For our calculations of $U/A$ and $\mu_s$ only, we used the value of $a = 3.42$ \AA, which is the average of 3.35 \AA\,\citep{lebedeva2011} and 3.495 \AA\,\citep{mostaani2015}, up to three significant digits.     

To obtain the wavenumber $k$, we choose one periodic path along the short diagonals of adjacent hexagons within the graphene layer, which we set as the $y$-axis, as displayed in Fig.~\ref{grapheneMultilayer}. However, this choice of the $y$-axis produces a double periodicity along the $x$-axis, which we do not address in our simplistic approximations of the graphene surface. Using the bond length of $1.42$ \AA, the short diagonal length is $\lambda = 2.46$ \AA\,\citep{chung2002}. This gives $k = 2\pi/(2.46 \mbox{\AA}) = 2.55$ \AA$^{-1}$. 

For graphene, the corrugation heights $h_1 = h_2 = h$ are associated with the atomic radii along the periodic path of corrugation, giving $h$ as the radius of a carbon atom, which is $0.914$ \AA\,\citep{pauling1947}.

Another factor that must be taken into account is the stacking of the graphene layers, with the layer shift being incorporated via the variable $y$. In the interest of simplicity, for calculating the interaction energy we consider the $AA$-stacked graphene configuration, corresponding to $y = 0$. It is important to note that the change in interaction energy with respect to $y$ is negligible in our model due to the dominance of the first three interaction energy terms in Eq.~(\ref{finalU}). 

Using these values, we calculated the interaction energy, maximum lateral force, normal force, and coefficient of static friction between two graphene layers as,
\begin{eqnarray}
\frac{U}{A} &=& -10.8 \frac{\mbox{meV}}{\mbox{atom}},\\
\frac{F_{L,\text{max}}}{A} &=& 6.46 \times 10^{-13} \frac{\mbox{N}}{\mbox{atom}},\\
\frac{F_N}{A} &=& -1.23 \times 10^{-11} \frac{\mbox{N}}{\mbox{atom}},\\
\mu_s &=& 0.053.
\end{eqnarray}
\subsection{Hexagonal Boron Nitride}
\begin{figure}[h]
    \centering
    \includegraphics[width=8.6cm]{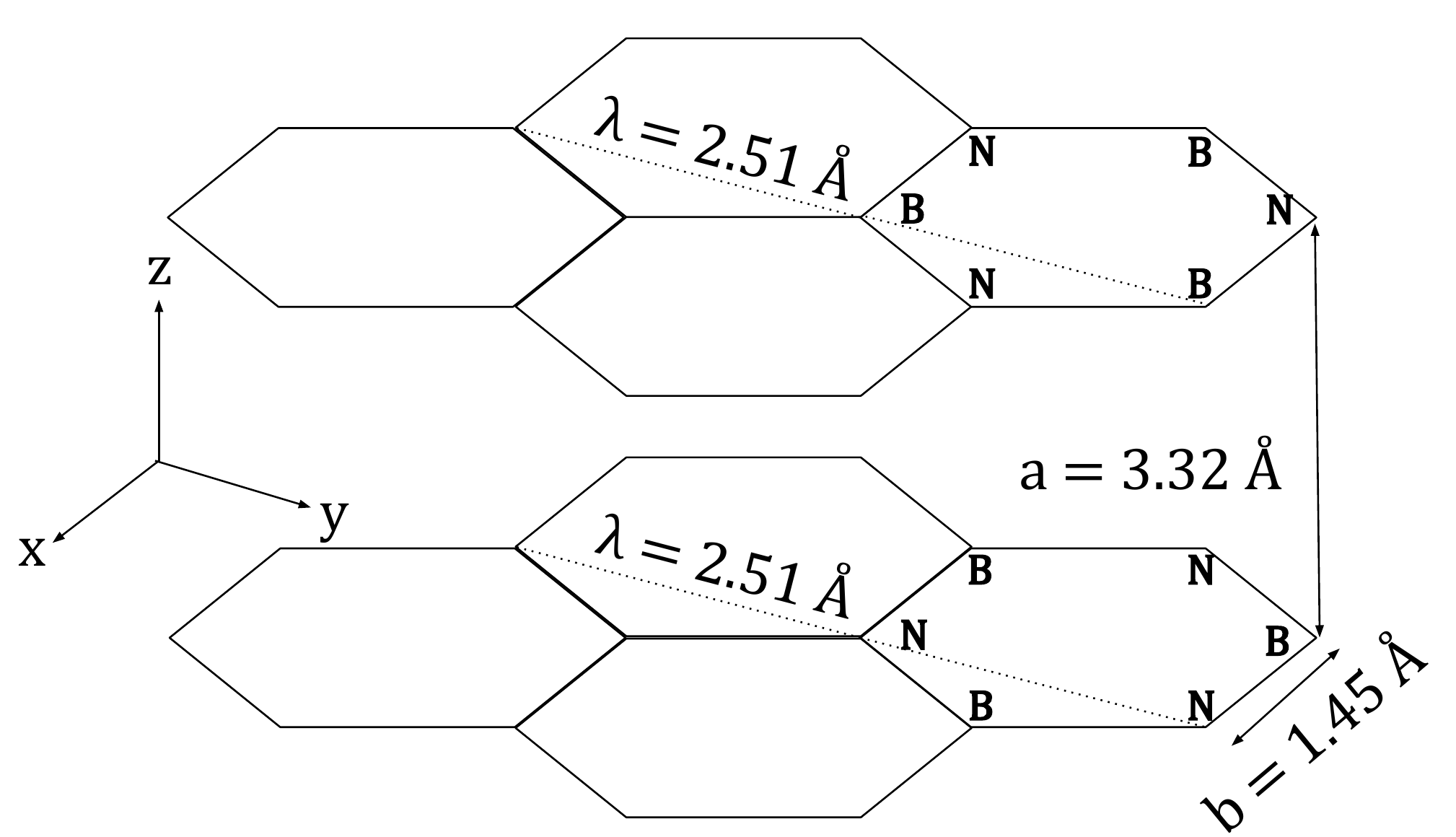}
    \caption{Hexagonal boron nitride, with the corrugation path being along the short diagonal. Note that atoms along a short diagonal will be the same, and atoms directly above each other in $AA^{\prime}$-stacked boron nitride will be opposite.}
    \label{hBNMultilayer}
\end{figure}

\begin{figure}[h]
    \centering
    \includegraphics[width=8.6cm]{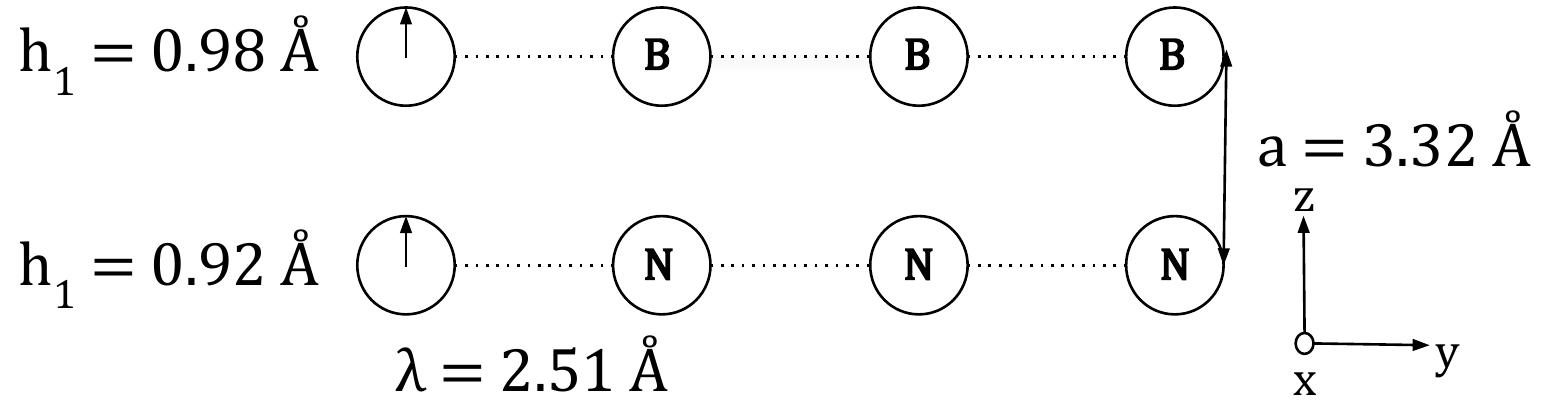}
    \caption{Interacting atoms along two vertically adjacent short diagonal paths, in $AA^{\prime}$-stacked h-BN. The atomic radii are treated as the corrugation heights. }
    \label{hBNInteraction}
\end{figure}

Hexagonal boron nitride (h-BN) has a similar hexagonal structure to graphene, with the fundamental difference being the alternation of adjacent atoms between nitrogen and boron. Figures~\ref{hBNMultilayer} and~\ref{hBNInteraction} illustrate the values of $a, h_1, h_2,$ and $\lambda$ in the structure of $AA'$-stacked h-BN.

The polarizability per unit cell of boron nitride is reported as 0.894 ${\text{\AA}}\,^3$\,\citep{kumar2016}. Using the unit cell area of 5.46 ${\text{\AA}}\,^2$\,\citep{hod2012}, the polarizability per unit area $\sigma$, will be $0.164$ \AA. 

For the interlayer distance $a$, we use $a =  3.32$\,\AA, which is the average of the two values $a = 3.30$ \AA\,\citep{shi2010}\,and $a = 3.33$ \AA\,\citep{solozhenko1995} reported in the literature, up to three significant digits. 

We use the radius of a boron atom for $h_1$ and the radius of a nitrogen atom for $h_2$. As there is both boron and nitrogen in each layer, this would approximate the interaction between two layers of h-BN to the interaction between a layer of boron and a layer of nitrogen. However, vertically adjacent atoms in $AA^{\prime}$-stacked boron nitride must alternate, and thus the interaction between alternate atoms should dominate. This approximation does not address the presence of both boron and nitrogen atoms within a single layer, and an improved treatment of such structures may be used in future iterations of this work. We have $h_1 = 0.98$ \AA\,and $h_2 = 0.92$ \AA\,\citep{pauling1947}. 

The corrugation wavelength, $\lambda$, of an h-BN surface is once again the short diagonal length. So, $\lambda = 2.51$ \AA, and $k = \frac{2\pi}{\lambda} = 2.50$ ${\text{\AA}}\,^{-1}$. 

For calculating the interaction energy, we use the $AA^{\prime}$-stacking configuration of h-BN in the interest of simplicity, with $y = 0$ \AA.

Using these values, we calculate the interaction energy, maximum lateral force, normal force, and coefficient of static friction between h-BN layers. Summarized, our results for hexagonal boron nitride are
\begin{eqnarray}
\frac{U}{A} &=& -11.4 \frac{\mbox{meV}}{\mbox{atom}},\\
\frac{F_{L,\text{max}}}{A} &=& 9.72 \times 10^{-13} \frac{\mbox{N}}{\mbox{atom}},\\
\frac{F_N}{A} &=& -1.27 \times 10^{-11} \frac{\mbox{N}}{\mbox{atom}},\\
\mu_s &=& 0.076.
\end{eqnarray}
\subsection{Black Phosphorene}
\begin{figure}[h]
    \centering
    \includegraphics[width=8.6cm]{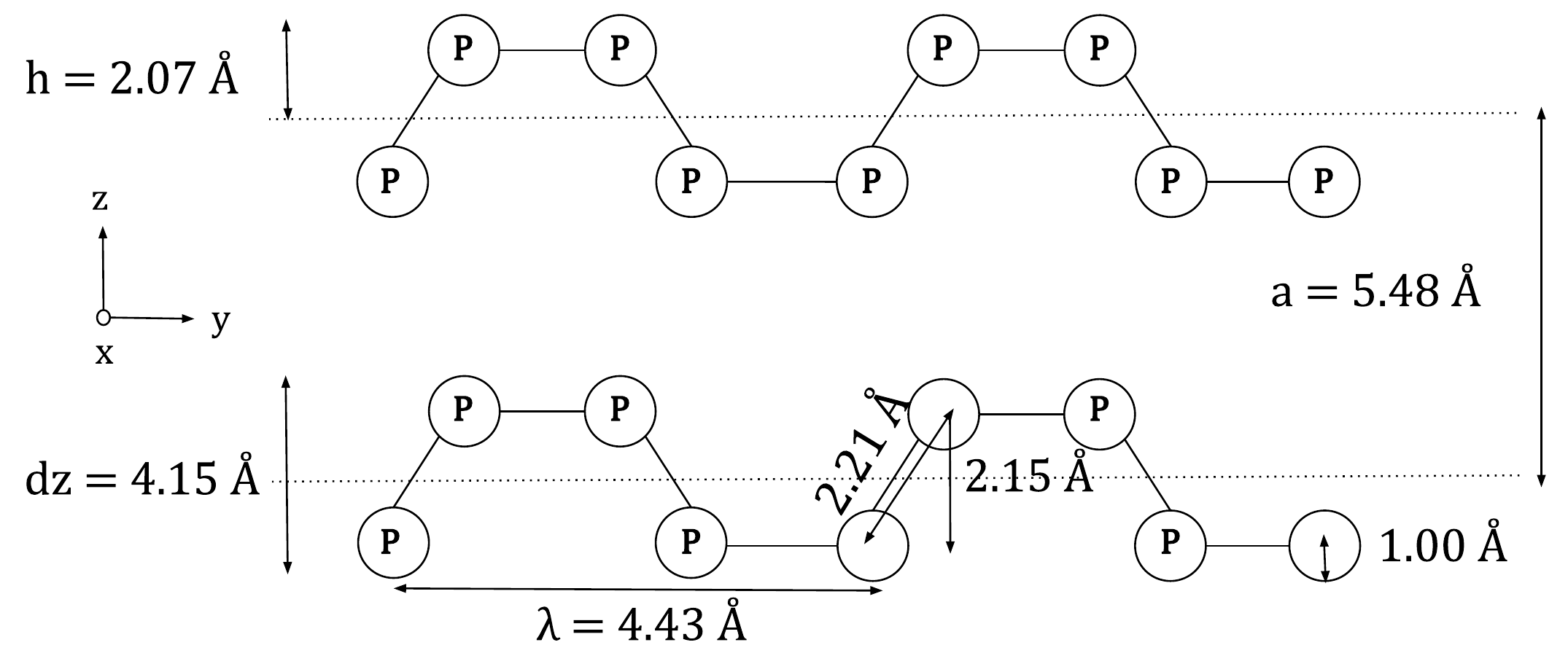}
    \caption{Profile of black phosphorene layer interaction. The height of the corrugation is now associated with the change in height over the long bond, which incorporates the atomic radius as well. The dotted lines through the center of each layer represent the midlines of corrugation.}
    \label{blackPhosphorus}
\end{figure}

Black phosphorene exhibits anisotropy with an armchair direction, which features a jagged periodic profile, and a zigzag direction, which is perpendicular to the armchair direction. We set the $y$-axis along the armchair direction. Figure~\ref{blackPhosphorus} demonstrates the black phosphorene corrugation structure in this direction.

A rectangular unit cell of black phosphorene, containing 4 atoms, has dimensions of 4.43 \AA $\times$ 3.28 \AA~\citep{jain2015}. The area per atom therefore is $(4.43 \mbox{\AA})(3.28 \mbox{\AA})/4 = 3.63\,{\text{\AA}\,^2}$. As there is no value in the literature for the polarizability ($\alpha$) of phosphorene, we instead calculate $\sigma$ using the dielectric constant and Eq.~(\ref{sigmadielectric}).

The relative permittivity $(\varepsilon/\varepsilon_0)$ of monolayer black phosphorus is reported as 5.65 (Ref.~\citep{Debu2018} and reference therein). As shown by the diagram, the total thickness of the layer will be $dz = 2.15\,\mbox{\AA}\,+2(1.00\,\mbox{\AA}) = 4.15\,\mbox{\AA}$ \cite{slater1964, castellanosgomez2014}.  From this, we calculate $\sigma = 1.54$\,\AA. 

We use an interlayer distance of $a = 5.48$ \AA, which is the average of $a = 5.46$ \AA\,and $a = 5.49$\,\AA, found in the literature \citep{castellanosgomez2014}, up to three significant digits.

The corrugation heights $h_1 = h_2 = h$ are equal to half of the vertical distance between the centers of adjacent phosphorus atoms connected by a long bond, added to the radius of a phosphorus atom. Using the bond length, 2.205 \AA, and the bond angle, 103.69$\degree$\,\citep{castellanosgomez2014}, we obtain $2.15$ \AA\, for the vertical distance of the bond length, giving $1.07\,\mbox{\AA} + 1\,\mbox{\AA} = 2.07$ \AA\,for $h$ \cite{slater1964}. 

For black phosphorene, the corrugation wavelength $\lambda$ is $4.43$ \AA, so $k = 2\pi/(4.43\,\mbox{\AA}) = 1.42\, {\text{\AA}}\,^{-1}$\,\citep{castellanosgomez2014}. The values of $a$, $h$, and $\lambda$ are visually represented in Fig.~\ref{blackPhosphorus}. We consider the $AA$-stacking configuration with $y = 0$ \AA\,in the interest of simplicity.
Our results for the black phosphorene interaction energy, maximum lateral force, normal force, and coefficient of static friction are summarized as,
\begin{eqnarray}
\frac{U}{A} &=& -152
\frac{\mbox{meV}}{\mbox{atom}},\\
\frac{F_{L,\text{max}}}{A} &=& 1.26 \times 10^{-11} \frac{\mbox{N}}{\mbox{atom}},\\
\frac{F_N}{A} &=& -7.32 \times 10^{-11} \frac{\mbox{N}}{\mbox{atom}},\\
\mu_s &=& 0.171.
\end{eqnarray}
A summary of our results is shown in Table~\ref{table}, with comparison to literature values:

\begin{center}
\begin{table*}[t]
\begin{tabular}{|c|c|c|c|c|}
\hline
Material & Calculated Interaction Energy \footnote{These values have an associated error of 15 to 20 percent as discussed in Section.~\ref{framework}.\hspace{5.3cm}~} & Reported Interaction Energy & Calculated $\mu_s$ & Reported $\mu_s$\\
\hline
Graphene ($AA$) & $-10.8$ meV/atom & \makecell{$-11.5 \pm 0.9$ meV/atom \citep{mostaani2015}\\ $-10.4$ meV/atom \cite{lebedeva2011}} & $0.053$ & \makecell{0.07 \\ \citep{mandelli2017}}\\
\hline
Boron Nitride ($AA^{\prime}$) & $-11.4$  meV/atom & \makecell{$-14$ to $-38$ meV/atom \citep{Hsing2014, rydberg2003, marom2010}} & $0.076$ & \makecell{0.07 \\ \citep{gangopadhyay1997}}\\
\hline
\makecell{Black Phosphorene \\ (AA)} & $-152$ meV/atom  & \makecell{$-151$ meV/atom \citep{sansone2016} \\ $-83$ to $-127$ meV/atom \citep{dong2016}} & $0.171$ & \makecell{- 
} \\
\hline
\end{tabular}
\caption{Comparison of calculated values to values given in the literature.}
\label{table}%
\end{table*}%
\end{center}%
\section{Discussion}\label{discussion}
Previous works have sought to understand the force of static friction in a more fundamental way, primarily through the approaches of contact mechanics and asperity adhesion \cite{tabor1981, johnson1971, derjaguin1975}. In this work, we seek to address this problem with the hypothesis that the macroscopic static friction force originates from fundamental forces. In a simple test case of van der Waals materials, we attribute the static friction force acting between two layers to the long-range van der Waals forces existing between the layers. Additionally, we used several approximations including the simplistic sinusoidal corrugations model, the Casimir-Polder limit of the van der Waals interaction, and the perturbative expansion in $h_i/a$ for the calculation of the interaction energy. These approximations were made primarily in the interest of simplifying our calculations and could be more rigorously treated in further iterations of this research. In the expression for total interaction energy per unit area, given in Eq.~(\ref{finalU}), the first-order terms $U^{(1, 0)}$ and $U^{(0, 1)}$ evaluate to zero and do not contribute to the total interaction energy. 
The remaining contribution to the interaction energy between corrugated plates is thus entirely composed of zeroth and second order terms in $h_i/a$.
The perturbative interaction energy between uniformly charged corrugated plates will show similar behavior.
The expression for $\mu_s$ is the ratio of the maximum lateral force to the normal force, both of which are derived by taking the partial derivatives of the interaction energy given by Eq.~(\ref{finalU}). In Table \ref{table} Columns 2, 3, we show comparisons of calculated and reported theoretical values of the interaction energies between layers of graphene, h-BN, and black phosphorene.


For graphene, we determined the interaction energy between two $AA$-stacked layers to be $-10.8$ meV/atom. Two theoretical studies, the first using quantum Monte Carlo simulations and the second using van der Waals density functional calculations, have computed the interlayer binding energy of bilayer graphene to be $11.5 \pm 0.9$ meV/atom \cite{mostaani2015} and $10.4$ meV/atom \cite{lebedeva2011}, respectively. Our value of $-10.8$ meV/atom is in good agreement with these results. Through experimental measurement of the elastic deformation energy of graphite flakes, Liu et al. found the binding energy of highly oriented pyrolytic graphite stacks to be $31 \pm 2$ meV/atom \cite{liu2012}. The binding energy of bulk graphite is considerably larger than the interlayer energy of bilayer graphene, as the bulk binding energy incorporates interactions between a second adjacent layer and additional layers at further distances \cite{marom2010}. 

In the case of h-BN, we calculated the interaction energy between two $AA'$-stacked layers as $-11.4$ meV/atom.
Using density functional methods, a range of about $14$ to $38$ meV/atom for the binding energy of bilayer h-BN for interlayer distances ranging from $3.3$ \AA\,to $3.6$ \AA\,has been obtained. \cite{Hsing2014, rydberg2003, marom2010}
Our result of $-11.4$ meV/atom is of the same order as these theoretical values, and to the best of our knowledge, the interaction energy between two h-BN layers has not yet been experimentally determined.

While graphene and h-BN have different geometric ($h_i, a, k$) and dielectric parameters ($\sigma_i$), we observe that our calculated interaction energies for these materials are close ($-10.8$ meV/atom and $-11.4$ meV/atom, respectively, see Table \ref{table} Column 2). This could be attributed to the greater atomic polarizability of graphene being offset by its greater interlayer distance, making the interlayer interaction energies similar. Further, while graphene and h-BN both have hexagonal structures with similar bond lengths, the surface interactions between layers of these materials differ due to the polar covalent B-N bond, resulting in partial charges near the boron and nitrogen atomic centers \cite{hod2012}. This results in additional electrostatic interactions between h-BN layers which are not present in the interaction between graphene layers \cite{hod2012}. However, Hod calculated the electrostatic interaction energy between $AA'$-stacked layers of h-BN and found the contributions of electrostatic effects to be negligible \cite{hod2012}.

The interaction energy between two layers of black phosphorene determined using our expression is $-152$ meV/atom. Using density functional theoretical calculations, Sansone et al. determined the exfoliation energy of black phosphorus to be $151$ meV/atom. Our result of $-152$ meV/atom is in good agreement with this theoretical value \cite{sansone2016}. Dong et al. performed van der Waals density functional (vdW DF) calculations using several different methods and determined the interlayer coupling of black phosphorus to be in the range of $83$ to $127$ meV/atom, which is of the same order as our result \cite{dong2016}. 

Our research shows that the coefficient of static friction $\mu_s$ is completely determined by geometric properties of a van der Waals material, particularly the interlayer distance and the amplitude and wavelength of the corrugations. In addition, the coefficient of static friction is explicitly independent of dielectric properties regardless of whether the normal force is approximated to the dominant term of Eq.~(\ref{normal1}) or not.  It should however be noted that in van der Waals materials, the interlayer distance $a$ is mediated by the strength of the van der Waals interaction, which in turn is dependent on dielectric properties \cite{marom2010}. The theoretical expression for $\mu_s$ may be calculated for van der Waals materials using only the lattice constants of the material and the corrugation height. We used the atomic radii as the corrugation heights; however, alternative approaches to determining the corrugation heights of the layers may be considered. Based on our expression, $\mu_s$ appears to be independent of the stacking configurations of van der Waals materials, which are incorporated into our model via the lateral shift $y$. However, this independence only holds when the normal force is approximated to the dominant contribution; if the $\cos(ky)$ term of the normal force was incorporated, $\mu_s$ would be dependent on the lateral shift and consequently the stacking configuration. Note that the dominance of the first three normal force terms implies that $\mu_s$ would not vary significantly with respect to $y$ if additional normal force terms were incorporated into Eq.~(\ref{equationmus}). In addition, Eq.~(\ref{equationmus}) and Fig.~(\ref{mus1}) demonstrate that $\mu_s$ approaches zero as $ka \to 0$ or $ka \to \infty$, where $k = 2\pi/\lambda$. This suggests a role of ``smoothness" in minimizing $\mu_s$, with $\lambda \gg a$ implying a nearly flat surface and $\lambda \ll a$ implying a surface with small enough corrugation wavelength to imitate a flat surface with only the adjacent corrugation peaks. Considering Eq.~(\ref{equationmus}), if one assumes the conservative bound of $h_1 \leq a/2$ and $h_2 \leq a/2$, the coefficient of static friction would have a maximum value of 1.29. Previously reported $\mu_s$ values for several van der Waals materials, including those discussed in this work, have not exceeded 1.29 \cite{mandelli2017, gangopadhyay1997, gradt2016, united1976friction}. Thus, under the assumption that layers in van der Waals materials do not geometrically lock into each other like fingers in a glove ($h_1 \leq a/2$ and $h_2 \leq a/2$), we can predict a theoretical bound of 1.29 for the coefficient of static friction between the layers.

Our expression for $\mu_s$ predicts the coefficient of static friction of van der Waals materials to be in reasonable agreement with previous results. Using this expression for $\mu_s$, the coefficient of static friction between two graphene layers is $0.053$. Mandelli et al. performed molecular dynamics simulations to determine the coefficient of static friction between two graphene layers, calculating $\mu_s$ to be 0.06 for small flake sizes and 0.07 in the limit of large flake sizes \cite{mandelli2017}. Assuming that our model of plates extending to infinity is closer to the limit of large flake size, we find reasonable agreement between our theoretical value and the reported computational value (see Table \ref{table} Columns 4, 5). The value of $\mu_s$ between h-BN layers is found to be $0.076$. Using a pins-on-rings tribometer experiment, Gangopadhyay et al. determined the coefficient of static friction between h-BN layers to be 0.07 \cite{gangopadhyay1997}, which is in good agreement with the calculated value. In the case of black phosphorus, we determined the value of $\mu_s$ between layers of black phosphorene to be $0.171$. As the study of black phosphorene is an emerging field of research, we were unable to find a value for the coefficient of static friction between two black phosphorene layers in the literature. The value we have derived using our expression may be of use for verification in future experiments involving static friction between layers of black phosphorene.
\section{Conclusions}\label{conclusions}
It is remarkable that our simplistic model, with the various assumptions described above, is able to determine the coefficient of static friction in reasonable agreement with previous results for these test cases of van der Waals materials. In addition, our predicted maximum of 1.29 for the coefficient of static of friction between van der Waals material layers appears to be supported by the literature. However, the approximations that we used may produce limitations of the model that could be overcome in future work. For example, we modeled the infinite plates as having one-dimensional $y$-direction sinusoidal corrugations to model the in the interest of simplicity. More complex surface roughness profiles could be better represented with a Fourier series involving both $x$- and $y$-direction cosine terms instead of just $y$-direction terms for the surface corrugations, as defined in Eq.~(\ref{corrugationsnew}). It should be noted that other authors have developed theoretical frameworks for treating surface roughness profiles in the context of Casimir interaction energy calculations \cite{wu2014, Broer2012}.

Although our current model was developed for calculating $\mu_s$ using Casimir-Polder interactions, it could be generalized to include the frequency-dependent contributions of van der Waals interactions and electrostatic interactions. A generalized model would elucidate the contributions of various fundamental interactions to the static friction force between two surfaces, while also describing the role of atomic-level properties in determining $\mu_s$. 

Our results may have potential uses in nanoengineering, because our analytic expression for $\mu_s$ could serve as a guide in the manipulation of static friction between van der Waals material layers. For example, the introduction of monolayer coatings in microelectromechanical systems has been proposed in order to reduce static friction at surface contacts \cite{maboudian2000}. As shown by Eq.~(\ref{equationmus}) and Fig.~\ref{mus1}, the coefficient of static friction between two layers is reduced for $h_1 \ll a$, $h_2 \ll a$, and for either $a \ll \lambda/2\pi$ or $a \gg \lambda/2\pi$. Our expression for $\mu_s$ would thus be useful in engineering such coatings to minimize $\mu_s$.    
\section*{Acknowledgements}
We would like to acknowledge John A. Logan College and Southern Illinois University--Carbondale for hospitality. In addition, we thank Prof. Punit Kohli, Prof. Saikat Talapatra, and Ms. Melissa Doellman for valuable discussions and comments.
\section*{Appendix}
\appendix
\section{Normal Force}
The normal force between corrugated plates, with the complete second-order terms in $h/a$ along with the zeroth order term given in Eq.~(\ref{normal1}), is
\begin{eqnarray}
\frac{F_N}{A} &=& -\frac{46\sigma_1\sigma_2\hbar c \pi}{(4\pi)^3a^6}\Big[1 + \frac{21}{2}\frac{h^2_1 + h^2_2}{a^2}\nonumber\\
&& +\frac{h_1}{a}\frac{h_2}{a}\frac{e^{-ka}}{30}Q(ka)\Big]\label{normalfull}
\end{eqnarray}
where for brevity we have defined 
\begin{eqnarray}
Q(ka) &=& (ka)^5 + 12(ka)^4+ 75(ka)^3\nonumber\\
&+& 285(ka)^2 + 630(ka) + 630.
\end{eqnarray}
\bibliography{biblio/friction}

\begin{thebibliography}{43}%
\makeatletter
\providecommand \@ifxundefined [1]{%
 \@ifx{#1\undefined}
}%
\providecommand \@ifnum [1]{%
 \ifnum #1\expandafter \@firstoftwo
 \else \expandafter \@secondoftwo
 \fi
}%
\providecommand \@ifx [1]{%
 \ifx #1\expandafter \@firstoftwo
 \else \expandafter \@secondoftwo
 \fi
}%
\providecommand \natexlab [1]{#1}%
\providecommand \enquote  [1]{``#1''}%
\providecommand \bibnamefont  [1]{#1}%
\providecommand \bibfnamefont [1]{#1}%
\providecommand \citenamefont [1]{#1}%
\providecommand \href@noop [0]{\@secondoftwo}%
\providecommand \href [0]{\begingroup \@sanitize@url \@href}%
\providecommand \@href[1]{\@@startlink{#1}\@@href}%
\providecommand \@@href[1]{\endgroup#1\@@endlink}%
\providecommand \@sanitize@url [0]{\catcode `\\12\catcode `\$12\catcode
  `\&12\catcode `\#12\catcode `\^12\catcode `\_12\catcode `\%12\relax}%
\providecommand \@@startlink[1]{}%
\providecommand \@@endlink[0]{}%
\providecommand \url  [0]{\begingroup\@sanitize@url \@url }%
\providecommand \@url [1]{\endgroup\@href {#1}{\urlprefix }}%
\providecommand \urlprefix  [0]{URL }%
\providecommand \Eprint [0]{\href }%
\providecommand \doibase [0]{http://dx.doi.org/}%
\providecommand \selectlanguage [0]{\@gobble}%
\providecommand \bibinfo  [0]{\@secondoftwo}%
\providecommand \bibfield  [0]{\@secondoftwo}%
\providecommand \translation [1]{[#1]}%
\providecommand \BibitemOpen [0]{}%
\providecommand \bibitemStop [0]{}%
\providecommand \bibitemNoStop [0]{.\EOS\space}%
\providecommand \EOS [0]{\spacefactor3000\relax}%
\providecommand \BibitemShut  [1]{\csname bibitem#1\endcsname}%
\let\auto@bib@innerbib\@empty
\bibitem [{\citenamefont {Popova}\ and\ \citenamefont
  {Popov}(2015)}]{popovfriction}%
  \BibitemOpen
  \bibfield  {author} {\bibinfo {author} {\bibfnamefont {E.}~\bibnamefont
  {Popova}}\ and\ \bibinfo {author} {\bibfnamefont {V.~L.}\ \bibnamefont
  {Popov}},\ }\bibfield  {title} {\enquote {\bibinfo {title} {{The Research
  Works of Coulomb and Amontons and Generalized Laws of Friction}},}\ }\href
  {\doibase 10.1007/s40544-015-0074-6} {\bibfield  {journal} {\bibinfo
  {journal} {Friction}\ }\textbf {\bibinfo {volume} {3}},\ \bibinfo {pages} {8}
  (\bibinfo {year} {2015})}\BibitemShut {NoStop}%
\bibitem [{\citenamefont {Dowson}(1978)}]{dowson1978}%
  \BibitemOpen
  \bibfield  {author} {\bibinfo {author} {\bibfnamefont {D.}~\bibnamefont
  {Dowson}},\ }\bibfield  {title} {\enquote {\bibinfo {title} {{{Men of
  Tribology: Charles Augustin Coulomb (1736–1806) and Arthur-Jules Morin
  (1795–1880)}}},}\ }\href {\doibase 10.1115/1.3453126} {\bibfield  {journal}
  {\bibinfo  {journal} {Journal of Lubrication Technology}\ }\textbf {\bibinfo
  {volume} {100}},\ \bibinfo {pages} {148--155} (\bibinfo {year}
  {1978})}\BibitemShut {NoStop}%
\bibitem [{\citenamefont {Blau}(2001)}]{blau2001}%
  \BibitemOpen
  \bibfield  {author} {\bibinfo {author} {\bibfnamefont {P.~J.}\ \bibnamefont
  {Blau}},\ }\bibfield  {title} {\enquote {\bibinfo {title} {{T}he
  {S}ignificance and {U}se of the {F}riction {C}oefficient},}\ }\href {\doibase
  https://doi.org/10.1016/S0301-679X(01)00050-0} {\bibfield  {journal}
  {\bibinfo  {journal} {Tribology International}\ }\textbf {\bibinfo {volume}
  {34}},\ \bibinfo {pages} {585--591} (\bibinfo {year} {2001})}\BibitemShut
  {NoStop}%
\bibitem [{\citenamefont {Bowden}\ \emph {et~al.}(1939)\citenamefont {Bowden},
  \citenamefont {Tabor},\ and\ \citenamefont {Taylor}}]{bowden1939}%
  \BibitemOpen
  \bibfield  {author} {\bibinfo {author} {\bibfnamefont {F.~P.}\ \bibnamefont
  {Bowden}}, \bibinfo {author} {\bibfnamefont {D.}~\bibnamefont {Tabor}}, \
  and\ \bibinfo {author} {\bibfnamefont {G.~I.}\ \bibnamefont {Taylor}},\
  }\bibfield  {title} {\enquote {\bibinfo {title} {{The Area of Contact between
  Stationary and Moving Surfaces}},}\ }\href {\doibase 10.1098/rspa.1939.0005}
  {\bibfield  {journal} {\bibinfo  {journal} {Proceedings of the Royal Society
  of London. Series A. Mathematical and Physical Sciences}\ }\textbf {\bibinfo
  {volume} {169}},\ \bibinfo {pages} {391--413} (\bibinfo {year}
  {1939})}\BibitemShut {NoStop}%
\bibitem [{\citenamefont {Tabor}(1981)}]{tabor1981}%
  \BibitemOpen
  \bibfield  {author} {\bibinfo {author} {\bibfnamefont {D.}~\bibnamefont
  {Tabor}},\ }\bibfield  {title} {\enquote {\bibinfo {title}
  {{{F}riction—{T}he {P}resent {S}tate of {O}ur {U}nderstanding}},}\ }\href
  {\doibase 10.1115/1.3251622} {\bibfield  {journal} {\bibinfo  {journal}
  {Journal of Lubrication Technology}\ }\textbf {\bibinfo {volume} {103}},\
  \bibinfo {pages} {169--179} (\bibinfo {year} {1981})}\BibitemShut {NoStop}%
\bibitem [{\citenamefont {Derjaguin}\ \emph {et~al.}(1975)\citenamefont
  {Derjaguin}, \citenamefont {Muller},\ and\ \citenamefont
  {Toporov}}]{derjaguin1975}%
  \BibitemOpen
  \bibfield  {author} {\bibinfo {author} {\bibfnamefont {B.~V.}\ \bibnamefont
  {Derjaguin}}, \bibinfo {author} {\bibfnamefont {V.~M.}\ \bibnamefont
  {Muller}}, \ and\ \bibinfo {author} {\bibfnamefont {Y.~P.}\ \bibnamefont
  {Toporov}},\ }\bibfield  {title} {\enquote {\bibinfo {title} {{{E}ffect of
  {C}ontact {D}eformations on the {A}dhesion of {P}articles}},}\ }\href
  {\doibase https://doi.org/10.1016/0021-9797(75)90018-1} {\bibfield  {journal}
  {\bibinfo  {journal} {Journal of Colloid and Interface Science}\ }\textbf
  {\bibinfo {volume} {53}},\ \bibinfo {pages} {314--326} (\bibinfo {year}
  {1975})}\BibitemShut {NoStop}%
\bibitem [{\citenamefont {Johnson}\ \emph {et~al.}(1971)\citenamefont
  {Johnson}, \citenamefont {Kendall}, \citenamefont {Roberts},\ and\
  \citenamefont {Tabor}}]{johnson1971}%
  \BibitemOpen
  \bibfield  {author} {\bibinfo {author} {\bibfnamefont {K.~L.}\ \bibnamefont
  {Johnson}}, \bibinfo {author} {\bibfnamefont {K.}~\bibnamefont {Kendall}},
  \bibinfo {author} {\bibfnamefont {A.~D.}\ \bibnamefont {Roberts}}, \ and\
  \bibinfo {author} {\bibfnamefont {D.}~\bibnamefont {Tabor}},\ }\bibfield
  {title} {\enquote {\bibinfo {title} {{Surface Energy and the Contact of
  Elastic Solids}},}\ }\href {\doibase 10.1098/rspa.1971.0141} {\bibfield
  {journal} {\bibinfo  {journal} {Proceedings of the Royal Society of London.
  A. Mathematical and Physical Sciences}\ }\textbf {\bibinfo {volume} {324}},\
  \bibinfo {pages} {301--313} (\bibinfo {year} {1971})}\BibitemShut {NoStop}%
\bibitem [{\citenamefont {Tabor}(1977)}]{tabor1977}%
  \BibitemOpen
  \bibfield  {author} {\bibinfo {author} {\bibfnamefont {D.}~\bibnamefont
  {Tabor}},\ }\bibfield  {title} {\enquote {\bibinfo {title} {{S}urface
  {F}orces and {S}urface {I}nteractions},}\ }\href {\doibase
  https://doi.org/10.1016/0021-9797(77)90366-6} {\bibfield  {journal} {\bibinfo
   {journal} {Journal of Colloid and Interface Science}\ }\textbf {\bibinfo
  {volume} {58}},\ \bibinfo {pages} {2--13} (\bibinfo {year} {1977})},\
  \bibinfo {note} {{I}nternational Conference on Colloids and
  Surfaces}\BibitemShut {NoStop}%
\bibitem [{\citenamefont {Mandelli}\ \emph {et~al.}(2017)\citenamefont
  {Mandelli}, \citenamefont {Leven}, \citenamefont {Hod},\ and\ \citenamefont
  {Urbakh}}]{mandelli2017}%
  \BibitemOpen
  \bibfield  {author} {\bibinfo {author} {\bibfnamefont {D.}~\bibnamefont
  {Mandelli}}, \bibinfo {author} {\bibfnamefont {I.}~\bibnamefont {Leven}},
  \bibinfo {author} {\bibfnamefont {O.}~\bibnamefont {Hod}}, \ and\ \bibinfo
  {author} {\bibfnamefont {M.}~\bibnamefont {Urbakh}},\ }\bibfield  {title}
  {\enquote {\bibinfo {title} {{Sliding Friction of Graphene/Hexagonal--Boron
  Nitride Heterojunctions: a Route to Robust Superlubricity}},}\ }\href
  {\doibase 10.1038/s41598-017-10522-8} {\bibfield  {journal} {\bibinfo
  {journal} {Scientific Reports}\ }\textbf {\bibinfo {volume} {7}},\ \bibinfo
  {pages} {10851} (\bibinfo {year} {2017})}\BibitemShut {NoStop}%
\bibitem [{\citenamefont {Jahangiri}\ \emph {et~al.}(2016)\citenamefont
  {Jahangiri}, \citenamefont {Heverly-Coulson},\ and\ \citenamefont
  {Mosey}}]{jahangiri2016}%
  \BibitemOpen
  \bibfield  {author} {\bibinfo {author} {\bibfnamefont {S.}~\bibnamefont
  {Jahangiri}}, \bibinfo {author} {\bibfnamefont {G.~S.}\ \bibnamefont
  {Heverly-Coulson}}, \ and\ \bibinfo {author} {\bibfnamefont {N.~J.}\
  \bibnamefont {Mosey}},\ }\bibfield  {title} {\enquote {\bibinfo {title}
  {{D}evelopment and {A}ssessment of {A}tomistic {M}odels for {P}redicting
  {S}tatic {F}riction {C}oefficients},}\ }\href {\doibase
  10.1103/PhysRevB.94.075406} {\bibfield  {journal} {\bibinfo  {journal} {Phys.
  Rev. B}\ }\textbf {\bibinfo {volume} {94}},\ \bibinfo {pages} {075406}
  (\bibinfo {year} {2016})}\BibitemShut {NoStop}%
\bibitem [{\citenamefont {M\"user}\ and\ \citenamefont
  {Wenning}(2001)}]{muser2001}%
  \BibitemOpen
  \bibfield  {author} {\bibinfo {author} {\bibfnamefont {M.~H.}\ \bibnamefont
  {M\"user}}\ and\ \bibinfo {author} {\bibfnamefont {M.~O}\ \bibnamefont
  {Wenning}, \bibfnamefont {L.~and~Robbins}},\ }\bibfield  {title} {\enquote
  {\bibinfo {title} {{S}imple {M}icroscopic {T}heory of {A}montons's {L}aws for
  {S}tatic {F}riction},}\ }\href {\doibase 10.1103/PhysRevLett.86.1295}
  {\bibfield  {journal} {\bibinfo  {journal} {Phys. Rev. Lett.}\ }\textbf
  {\bibinfo {volume} {86}},\ \bibinfo {pages} {1295--1298} (\bibinfo {year}
  {2001})}\BibitemShut {NoStop}%
\bibitem [{\citenamefont {Marom}\ \emph {et~al.}(2010)\citenamefont {Marom},
  \citenamefont {Bernstein}, \citenamefont {Garel}, \citenamefont {Tkatchenko},
  \citenamefont {Joselevich}, \citenamefont {Kronik},\ and\ \citenamefont
  {Hod}}]{marom2010}%
  \BibitemOpen
  \bibfield  {author} {\bibinfo {author} {\bibfnamefont {N.}~\bibnamefont
  {Marom}}, \bibinfo {author} {\bibfnamefont {J.}~\bibnamefont {Bernstein}},
  \bibinfo {author} {\bibfnamefont {J.}~\bibnamefont {Garel}}, \bibinfo
  {author} {\bibfnamefont {A.}~\bibnamefont {Tkatchenko}}, \bibinfo {author}
  {\bibfnamefont {E.}~\bibnamefont {Joselevich}}, \bibinfo {author}
  {\bibfnamefont {L.}~\bibnamefont {Kronik}}, \ and\ \bibinfo {author}
  {\bibfnamefont {O.}~\bibnamefont {Hod}},\ }\bibfield  {title} {\enquote
  {\bibinfo {title} {{Stacking and Registry Effects in Layered Materials: The
  Case of Hexagonal Boron Nitride}},}\ }\href {\doibase
  10.1103/PhysRevLett.105.046801} {\bibfield  {journal} {\bibinfo  {journal}
  {Phys. Rev. Lett.}\ }\textbf {\bibinfo {volume} {105}},\ \bibinfo {pages}
  {046801} (\bibinfo {year} {2010})}\BibitemShut {NoStop}%
\bibitem [{\citenamefont {Lifshitz}(1955)}]{lifshitz1956}%
  \BibitemOpen
  \bibfield  {author} {\bibinfo {author} {\bibfnamefont {E.~M.}\ \bibnamefont
  {Lifshitz}},\ }\bibfield  {title} {\enquote {\bibinfo {title} {{The Theory of
  Molecular Attractive Forces between Solids}},}\ }\href@noop {} {\bibfield
  {journal} {\bibinfo  {journal} {Pis’ma Zh. Eksp. Teor. Fiz.}\ }\textbf
  {\bibinfo {volume} {29}},\ \bibinfo {pages} {94} (\bibinfo {year} {1955})},\
  \bibinfo {note} {[English version:
  \href{http://www.jetp.ac.ru/cgi-bin/e/index/e/2/1/p73?a=list}{Sov. Phys. JETP
  {\bf 2}, 73 (1956)}]}\BibitemShut {NoStop}%
\bibitem [{\citenamefont {Dzyaloshinskii}\ \emph {et~al.}(1961)\citenamefont
  {Dzyaloshinskii}, \citenamefont {Lifshitz},\ and\ \citenamefont
  {Pitaevskii}}]{dzyaloshinskii1961}%
  \BibitemOpen
  \bibfield  {author} {\bibinfo {author} {\bibfnamefont {I.~E.}\ \bibnamefont
  {Dzyaloshinskii}}, \bibinfo {author} {\bibfnamefont {E.~M.}\ \bibnamefont
  {Lifshitz}}, \ and\ \bibinfo {author} {\bibfnamefont {L.~P.}\ \bibnamefont
  {Pitaevskii}},\ }\bibfield  {title} {\enquote {\bibinfo {title} {{General
  Theory of van der {W}aals' Forces}},}\ }\href {\doibase
  10.3367/UFNr.0073.196103b.0381} {\bibfield  {journal} {\bibinfo  {journal}
  {{Usp. Fiz. Nauk}}\ }\textbf {\bibinfo {volume} {73}},\ \bibinfo {pages}
  {381} (\bibinfo {year} {1961})},\ \bibinfo {note} {[English version:
  \href{https://doi.org/10.1070/PU1961v004n02ABEH003330}{Sov. Phys. Usp. {\bf
  4}, 153 (1961)}]}\BibitemShut {NoStop}%
\bibitem [{\citenamefont {Casimir}\ and\ \citenamefont
  {Polder}(1948)}]{casimir1948}%
  \BibitemOpen
  \bibfield  {author} {\bibinfo {author} {\bibfnamefont {H.~B.~G.}\
  \bibnamefont {Casimir}}\ and\ \bibinfo {author} {\bibfnamefont
  {D.}~\bibnamefont {Polder}},\ }\bibfield  {title} {\enquote {\bibinfo {title}
  {{The Influence of Retardation on the London-van der Waals Forces}},}\ }\href
  {\doibase 10.1103/PhysRev.73.360} {\bibfield  {journal} {\bibinfo  {journal}
  {Phys. Rev.}\ }\textbf {\bibinfo {volume} {73}},\ \bibinfo {pages} {360--372}
  (\bibinfo {year} {1948})}\BibitemShut {NoStop}%
\bibitem [{\citenamefont {Garcion}\ \emph {et~al.}(2021)\citenamefont
  {Garcion}, \citenamefont {Fabre}, \citenamefont {Bricha}, \citenamefont
  {Perales}, \citenamefont {Scheel}, \citenamefont {Ducloy},\ and\
  \citenamefont {Dutier}}]{garcion2021}%
  \BibitemOpen
  \bibfield  {author} {\bibinfo {author} {\bibfnamefont {C.}~\bibnamefont
  {Garcion}}, \bibinfo {author} {\bibfnamefont {N.}~\bibnamefont {Fabre}},
  \bibinfo {author} {\bibfnamefont {H.}~\bibnamefont {Bricha}}, \bibinfo
  {author} {\bibfnamefont {F.}~\bibnamefont {Perales}}, \bibinfo {author}
  {\bibfnamefont {S.}~\bibnamefont {Scheel}}, \bibinfo {author} {\bibfnamefont
  {M.}~\bibnamefont {Ducloy}}, \ and\ \bibinfo {author} {\bibfnamefont
  {G.}~\bibnamefont {Dutier}},\ }\bibfield  {title} {\enquote {\bibinfo {title}
  {{I}ntermediate-{R}ange {C}asimir-{P}older {I}nteraction {P}robed by
  {H}igh-{O}rder {S}low {A}tom {D}iffraction},}\ }\href {\doibase
  10.1103/PhysRevLett.127.170402} {\bibfield  {journal} {\bibinfo  {journal}
  {Phys. Rev. Lett.}\ }\textbf {\bibinfo {volume} {127}},\ \bibinfo {pages}
  {170402} (\bibinfo {year} {2021})}\BibitemShut {NoStop}%
\bibitem [{\citenamefont {Shajesh}\ \emph {et~al.}(2017)\citenamefont
  {Shajesh}, \citenamefont {Parashar},\ and\ \citenamefont
  {Brevik}}]{shajesh2017}%
  \BibitemOpen
  \bibfield  {author} {\bibinfo {author} {\bibfnamefont {K.~V.}\ \bibnamefont
  {Shajesh}}, \bibinfo {author} {\bibfnamefont {P.}~\bibnamefont {Parashar}}, \
  and\ \bibinfo {author} {\bibfnamefont {I.}~\bibnamefont {Brevik}},\
  }\bibfield  {title} {\enquote {\bibinfo {title} {{Casimir–Polder Energy for
  Axially Symmetric Systems}},}\ }\href {\doibase 10.1016/j.aop.2017.10.008}
  {\bibfield  {journal} {\bibinfo  {journal} {Annals of Physics}\ }\textbf
  {\bibinfo {volume} {387}},\ \bibinfo {pages} {166–202} (\bibinfo {year}
  {2017})}\BibitemShut {NoStop}%
\bibitem [{\citenamefont {Cavero-Pel\'aez}\ \emph {et~al.}(2008)\citenamefont
  {Cavero-Pel\'aez}, \citenamefont {Milton}, \citenamefont {Parashar},\ and\
  \citenamefont {Shajesh}}]{noncontactgears1}%
  \BibitemOpen
  \bibfield  {author} {\bibinfo {author} {\bibfnamefont {I.}~\bibnamefont
  {Cavero-Pel\'aez}}, \bibinfo {author} {\bibfnamefont {K.~A.}\ \bibnamefont
  {Milton}}, \bibinfo {author} {\bibfnamefont {P.}~\bibnamefont {Parashar}}, \
  and\ \bibinfo {author} {\bibfnamefont {K.~V.}\ \bibnamefont {Shajesh}},\
  }\bibfield  {title} {\enquote {\bibinfo {title} {{Noncontact Gears. I.
  Next-to-Leading Order Contribution to the Lateral Casimir Force Between
  Corrugated Parallel Plates}},}\ }\href {\doibase 10.1103/PhysRevD.78.065018}
  {\bibfield  {journal} {\bibinfo  {journal} {Phys. Rev. D}\ }\textbf {\bibinfo
  {volume} {78}},\ \bibinfo {pages} {065018} (\bibinfo {year}
  {2008})}\BibitemShut {NoStop}%
\bibitem [{\citenamefont {{Gradshteyn, I.~S.~ and Ryzhik,
  I.~M.~}}(2007)}]{gradshteyn2007}%
  \BibitemOpen
  \bibfield  {author} {\bibinfo {author} {\bibnamefont {{Gradshteyn, I.~S.~ and
  Ryzhik, I.~M.~}}},\ }\href@noop {} {\emph {\bibinfo {title} {{Table of
  {I}ntegrals, {S}eries, and {P}roducts}}}},\ \bibinfo {edition} {{S}eventh}\
  ed.\ (\bibinfo  {publisher} {Elsevier/Academic Press, Amsterdam},\ \bibinfo
  {year} {2007})\ \bibinfo {note} {3.773 No. 6, 6.596 No. 3}\BibitemShut
  {NoStop}%
\bibitem [{\citenamefont {Zhong}\ and\ \citenamefont
  {Tom\'anek}(1990)}]{zhong1990}%
  \BibitemOpen
  \bibfield  {author} {\bibinfo {author} {\bibfnamefont {W.}~\bibnamefont
  {Zhong}}\ and\ \bibinfo {author} {\bibfnamefont {D.}~\bibnamefont
  {Tom\'anek}},\ }\bibfield  {title} {\enquote {\bibinfo {title}
  {{F}irst-{P}rinciples {T}heory of {A}tomic-{S}cale {F}riction},}\ }\href
  {\doibase 10.1103/PhysRevLett.64.3054} {\bibfield  {journal} {\bibinfo
  {journal} {Phys. Rev. Lett.}\ }\textbf {\bibinfo {volume} {64}},\ \bibinfo
  {pages} {3054--3057} (\bibinfo {year} {1990})}\BibitemShut {NoStop}%
\bibitem [{\citenamefont {Kumar}\ \emph {et~al.}(2016)\citenamefont {Kumar},
  \citenamefont {Chauhan}, \citenamefont {Agarwal},\ and\ \citenamefont
  {Bhowmick}}]{kumar2016}%
  \BibitemOpen
  \bibfield  {author} {\bibinfo {author} {\bibfnamefont {P.}~\bibnamefont
  {Kumar}}, \bibinfo {author} {\bibfnamefont {Y.~S.}\ \bibnamefont {Chauhan}},
  \bibinfo {author} {\bibfnamefont {A.}~\bibnamefont {Agarwal}}, \ and\
  \bibinfo {author} {\bibfnamefont {S.}~\bibnamefont {Bhowmick}},\ }\bibfield
  {title} {\enquote {\bibinfo {title} {{Thickness and Stacking Dependent
  Polarizability and Dielectric Constant of Graphene–Hexagonal Boron Nitride
  Composite Stacks}},}\ }\href {\doibase 10.1021/acs.jpcc.6b05805} {\bibfield
  {journal} {\bibinfo  {journal} {The Journal of Physical Chemistry C}\
  }\textbf {\bibinfo {volume} {120}},\ \bibinfo {pages} {17620--17626}
  (\bibinfo {year} {2016})}\BibitemShut {NoStop}%
\bibitem [{\citenamefont {Chung}(2002)}]{chung2002}%
  \BibitemOpen
  \bibfield  {author} {\bibinfo {author} {\bibfnamefont {D.~D.~L.}\
  \bibnamefont {Chung}},\ }\bibfield  {title} {\enquote {\bibinfo {title}
  {{Review Graphite}},}\ }\href {\doibase 10.1023/A:1014915307738} {\bibfield
  {journal} {\bibinfo  {journal} {Journal of Materials Science}\ }\textbf
  {\bibinfo {volume} {37}},\ \bibinfo {pages} {1475--1489} (\bibinfo {year}
  {2002})}\BibitemShut {NoStop}%
\bibitem [{\citenamefont {Lebedeva}\ \emph {et~al.}(2011)\citenamefont
  {Lebedeva}, \citenamefont {Knizhnik}, \citenamefont {Popov}, \citenamefont
  {Lozovik},\ and\ \citenamefont {Potapkin}}]{lebedeva2011}%
  \BibitemOpen
  \bibfield  {author} {\bibinfo {author} {\bibfnamefont {I.~V.}\ \bibnamefont
  {Lebedeva}}, \bibinfo {author} {\bibfnamefont {A.~A.}\ \bibnamefont
  {Knizhnik}}, \bibinfo {author} {\bibfnamefont {A.~M.}\ \bibnamefont {Popov}},
  \bibinfo {author} {\bibfnamefont {Y.~E.}\ \bibnamefont {Lozovik}}, \ and\
  \bibinfo {author} {\bibfnamefont {B.~V.}\ \bibnamefont {Potapkin}},\
  }\bibfield  {title} {\enquote {\bibinfo {title} {{{Interlayer Interaction and
  Relative Vibrations of Bilayer Graphene}}},}\ }\href {\doibase
  10.1039/C0CP02614J} {\bibfield  {journal} {\bibinfo  {journal} {Phys. Chem.
  Chem. Phys.}\ }\textbf {\bibinfo {volume} {13}},\ \bibinfo {pages}
  {5687--5695} (\bibinfo {year} {2011})}\BibitemShut {NoStop}%
\bibitem [{\citenamefont {Mostaani}\ \emph {et~al.}(2015)\citenamefont
  {Mostaani}, \citenamefont {Drummond},\ and\ \citenamefont
  {Fal'ko}}]{mostaani2015}%
  \BibitemOpen
  \bibfield  {author} {\bibinfo {author} {\bibfnamefont {E.}~\bibnamefont
  {Mostaani}}, \bibinfo {author} {\bibfnamefont {N.~D.}\ \bibnamefont
  {Drummond}}, \ and\ \bibinfo {author} {\bibfnamefont {V.~I.}\ \bibnamefont
  {Fal'ko}},\ }\bibfield  {title} {\enquote {\bibinfo {title} {{Quantum Monte
  Carlo Calculation of the Binding Energy of Bilayer Graphene}},}\ }\href
  {\doibase 10.1103/PhysRevLett.115.115501} {\bibfield  {journal} {\bibinfo
  {journal} {Phys. Rev. Lett.}\ }\textbf {\bibinfo {volume} {115}},\ \bibinfo
  {pages} {115501} (\bibinfo {year} {2015})}\BibitemShut {NoStop}%
\bibitem [{\citenamefont {Pauling}(1947)}]{pauling1947}%
  \BibitemOpen
  \bibfield  {author} {\bibinfo {author} {\bibfnamefont {L.}~\bibnamefont
  {Pauling}},\ }\bibfield  {title} {\enquote {\bibinfo {title} {{A}tomic
  {R}adii and {I}nteratomic {D}istances in {M}etals},}\ }\href {\doibase
  10.1021/ja01195a024} {\bibfield  {journal} {\bibinfo  {journal} {Journal of
  the American Chemical Society}\ }\textbf {\bibinfo {volume} {69}},\ \bibinfo
  {pages} {542--553} (\bibinfo {year} {1947})}\BibitemShut {NoStop}%
\bibitem [{\citenamefont {Hod}(2012)}]{hod2012}%
  \BibitemOpen
  \bibfield  {author} {\bibinfo {author} {\bibfnamefont {O.}~\bibnamefont
  {Hod}},\ }\bibfield  {title} {\enquote {\bibinfo {title} {{Graphite and
  Hexagonal Boron-Nitride have the Same Interlayer Distance. Why?}}}\ }\href
  {\doibase 10.1021/ct200880m} {\bibfield  {journal} {\bibinfo  {journal}
  {Journal of Chemical Theory and Computation}\ }\textbf {\bibinfo {volume}
  {8}},\ \bibinfo {pages} {1360--1369} (\bibinfo {year} {2012})},\ \bibinfo
  {note} {pMID: 26596751}\BibitemShut {NoStop}%
\bibitem [{\citenamefont {Shi}\ \emph {et~al.}(2010)\citenamefont {Shi},
  \citenamefont {Hamsen}, \citenamefont {Jia}, \citenamefont {Kim},
  \citenamefont {Reina}, \citenamefont {Hofmann}, \citenamefont {Hsu},
  \citenamefont {Li}, \citenamefont {Juang}, \citenamefont {Dresselhaus},\ and\
  \citenamefont {Kong}}]{shi2010}%
  \BibitemOpen
  \bibfield  {author} {\bibinfo {author} {\bibfnamefont {Y.}~\bibnamefont
  {Shi}}, \bibinfo {author} {\bibfnamefont {C.}~\bibnamefont {Hamsen}},
  \bibinfo {author} {\bibfnamefont {X.}~\bibnamefont {Jia}}, \bibinfo {author}
  {\bibfnamefont {K.~K.}\ \bibnamefont {Kim}}, \bibinfo {author} {\bibfnamefont
  {A.}~\bibnamefont {Reina}}, \bibinfo {author} {\bibfnamefont
  {M.}~\bibnamefont {Hofmann}}, \bibinfo {author} {\bibfnamefont
  {K.}~\bibnamefont {Hsu}, \bibfnamefont {A.~L.~and~Zhang}}, \bibinfo {author}
  {\bibfnamefont {H.}~\bibnamefont {Li}}, \bibinfo {author} {\bibfnamefont
  {Z.}~\bibnamefont {Juang}}, \bibinfo {author} {\bibfnamefont
  {L.}~\bibnamefont {Dresselhaus}, \bibfnamefont {M.~S.~and~Li}}, \ and\
  \bibinfo {author} {\bibfnamefont {J.}~\bibnamefont {Kong}},\ }\bibfield
  {title} {\enquote {\bibinfo {title} {{Synthesis of Few-Layer Hexagonal Boron
  Nitride Thin Film by Chemical Vapor Deposition}},}\ }\href {\doibase
  10.1021/nl1023707} {\bibfield  {journal} {\bibinfo  {journal} {Nano Letters}\
  }\textbf {\bibinfo {volume} {10}},\ \bibinfo {pages} {4134--4139} (\bibinfo
  {year} {2010})}\BibitemShut {NoStop}%
\bibitem [{\citenamefont {Solozhenko}\ \emph {et~al.}(1995)\citenamefont
  {Solozhenko}, \citenamefont {Will},\ and\ \citenamefont
  {Elf}}]{solozhenko1995}%
  \BibitemOpen
  \bibfield  {author} {\bibinfo {author} {\bibfnamefont {V.~L.}\ \bibnamefont
  {Solozhenko}}, \bibinfo {author} {\bibfnamefont {G.}~\bibnamefont {Will}}, \
  and\ \bibinfo {author} {\bibfnamefont {F.}~\bibnamefont {Elf}},\ }\bibfield
  {title} {\enquote {\bibinfo {title} {{Isothermal compression of hexagonal
  graphite-like boron nitride up to 12 GPa}},}\ }\href {\doibase
  https://doi.org/10.1016/0038-1098(95)00381-9} {\bibfield  {journal} {\bibinfo
   {journal} {Solid State Communications}\ }\textbf {\bibinfo {volume} {96}},\
  \bibinfo {pages} {1--3} (\bibinfo {year} {1995})}\BibitemShut {NoStop}%
\bibitem [{\citenamefont {Jain}\ and\ \citenamefont
  {Mcgaughey}(2015)}]{jain2015}%
  \BibitemOpen
  \bibfield  {author} {\bibinfo {author} {\bibfnamefont {A.}~\bibnamefont
  {Jain}}\ and\ \bibinfo {author} {\bibfnamefont {A.}~\bibnamefont
  {Mcgaughey}},\ }\bibfield  {title} {\enquote {\bibinfo {title} {{Strongly
  Anisotropic In-Plane Thermal Transport in Single-Layer Black Phosphorene}},}\
  }\href {\doibase 10.1038/srep08501} {\bibfield  {journal} {\bibinfo
  {journal} {Scientific reports}\ }\textbf {\bibinfo {volume} {5}},\ \bibinfo
  {pages} {8501} (\bibinfo {year} {2015})}\BibitemShut {NoStop}%
\bibitem [{\citenamefont {Debu}\ \emph {et~al.}(2018)\citenamefont {Debu},
  \citenamefont {Bauman}, \citenamefont {French}, \citenamefont {Churchill},\
  and\ \citenamefont {Herzog}}]{Debu2018}%
  \BibitemOpen
  \bibfield  {author} {\bibinfo {author} {\bibfnamefont {D.~T.}\ \bibnamefont
  {Debu}}, \bibinfo {author} {\bibfnamefont {S.~J.}\ \bibnamefont {Bauman}},
  \bibinfo {author} {\bibfnamefont {D.}~\bibnamefont {French}}, \bibinfo
  {author} {\bibfnamefont {H.~O.~H.}\ \bibnamefont {Churchill}}, \ and\
  \bibinfo {author} {\bibfnamefont {J.~B.}\ \bibnamefont {Herzog}},\ }\bibfield
   {title} {\enquote {\bibinfo {title} {{Tuning Infrared Plasmon Resonance of
  Black Phosphorene Nanoribbon with a Dielectric Interface}},}\ }\href
  {\doibase 10.1038/s41598-018-21365-2} {\bibfield  {journal} {\bibinfo
  {journal} {Scientific Reports}\ }\textbf {\bibinfo {volume} {8}},\ \bibinfo
  {pages} {3224} (\bibinfo {year} {2018})}\BibitemShut {NoStop}%
\bibitem [{\citenamefont {Slater}(1964)}]{slater1964}%
  \BibitemOpen
  \bibfield  {author} {\bibinfo {author} {\bibfnamefont {J.~C.}\ \bibnamefont
  {Slater}},\ }\bibfield  {title} {\enquote {\bibinfo {title} {{A}tomic {R}adii
  in {C}rystals},}\ }\href {\doibase 10.1063/1.1725697} {\bibfield  {journal}
  {\bibinfo  {journal} {The Journal of Chemical Physics}\ }\textbf {\bibinfo
  {volume} {41}},\ \bibinfo {pages} {3199--3204} (\bibinfo {year}
  {1964})}\BibitemShut {NoStop}%
\bibitem [{\citenamefont {Castellanos-Gomez}\ \emph {et~al.}(2014)\citenamefont
  {Castellanos-Gomez}, \citenamefont {Vicarelli}, \citenamefont {Prada},
  \citenamefont {Island}, \citenamefont {Narasimha-Acharya}, \citenamefont
  {Blanter}, \citenamefont {Groenendijk}, \citenamefont {Buscema},
  \citenamefont {Steele}, \citenamefont {Alvarez}, \citenamefont {Zandbergen},
  \citenamefont {Palacios},\ and\ \citenamefont {Zant}}]{castellanosgomez2014}%
  \BibitemOpen
  \bibfield  {author} {\bibinfo {author} {\bibfnamefont {A.}~\bibnamefont
  {Castellanos-Gomez}}, \bibinfo {author} {\bibfnamefont {L.}~\bibnamefont
  {Vicarelli}}, \bibinfo {author} {\bibfnamefont {E.}~\bibnamefont {Prada}},
  \bibinfo {author} {\bibfnamefont {J.}~\bibnamefont {Island}}, \bibinfo
  {author} {\bibfnamefont {K.}~\bibnamefont {Narasimha-Acharya}}, \bibinfo
  {author} {\bibfnamefont {S.}~\bibnamefont {Blanter}}, \bibinfo {author}
  {\bibfnamefont {D.}~\bibnamefont {Groenendijk}}, \bibinfo {author}
  {\bibfnamefont {M.}~\bibnamefont {Buscema}}, \bibinfo {author} {\bibfnamefont
  {G.}~\bibnamefont {Steele}}, \bibinfo {author} {\bibfnamefont
  {J.}~\bibnamefont {Alvarez}}, \bibinfo {author} {\bibfnamefont
  {H.}~\bibnamefont {Zandbergen}}, \bibinfo {author} {\bibfnamefont
  {J.}~\bibnamefont {Palacios}}, \ and\ \bibinfo {author} {\bibfnamefont
  {H.}~\bibnamefont {Zant}},\ }\bibfield  {title} {\enquote {\bibinfo {title}
  {{I}solation and {C}haracterization of {F}ew-{L}ayer {B}lack {P}hosphorus},}\
  }\href {\doibase 10.1088/2053-1583/1/2/025001} {\bibfield  {journal}
  {\bibinfo  {journal} {2D Materials}\ }\textbf {\bibinfo {volume} {1}},\
  \bibinfo {pages} {025001} (\bibinfo {year} {2014})}\BibitemShut {NoStop}%
\bibitem [{\citenamefont {Hsing}\ \emph {et~al.}(2014)\citenamefont {Hsing},
  \citenamefont {Cheng}, \citenamefont {Chou}, \citenamefont {Chang},\ and\
  \citenamefont {Wei}}]{Hsing2014}%
  \BibitemOpen
  \bibfield  {author} {\bibinfo {author} {\bibfnamefont {C.}~\bibnamefont
  {Hsing}}, \bibinfo {author} {\bibfnamefont {C.}~\bibnamefont {Cheng}},
  \bibinfo {author} {\bibfnamefont {J.}~\bibnamefont {Chou}}, \bibinfo {author}
  {\bibfnamefont {C.}~\bibnamefont {Chang}}, \ and\ \bibinfo {author}
  {\bibfnamefont {C.}~\bibnamefont {Wei}},\ }\bibfield  {title} {\enquote
  {\bibinfo {title} {{Van der Waals Interaction in a Boron Nitride Bilayer}},}\
  }\href {\doibase 10.1088/1367-2630/16/11/113015} {\bibfield  {journal}
  {\bibinfo  {journal} {New Journal of Physics}\ }\textbf {\bibinfo {volume}
  {16}},\ \bibinfo {pages} {113015} (\bibinfo {year} {2014})}\BibitemShut
  {NoStop}%
\bibitem [{\citenamefont {Rydberg}\ \emph {et~al.}(2003)\citenamefont
  {Rydberg}, \citenamefont {Dion}, \citenamefont {Jacobson}, \citenamefont
  {Schr\"oder}, \citenamefont {Hyldgaard}, \citenamefont {Simak}, \citenamefont
  {Langreth},\ and\ \citenamefont {Lundqvist}}]{rydberg2003}%
  \BibitemOpen
  \bibfield  {author} {\bibinfo {author} {\bibfnamefont {H.}~\bibnamefont
  {Rydberg}}, \bibinfo {author} {\bibfnamefont {M.}~\bibnamefont {Dion}},
  \bibinfo {author} {\bibfnamefont {N.}~\bibnamefont {Jacobson}}, \bibinfo
  {author} {\bibfnamefont {E.}~\bibnamefont {Schr\"oder}}, \bibinfo {author}
  {\bibfnamefont {P.}~\bibnamefont {Hyldgaard}}, \bibinfo {author}
  {\bibfnamefont {S.~I.}\ \bibnamefont {Simak}}, \bibinfo {author}
  {\bibfnamefont {D.~C.}\ \bibnamefont {Langreth}}, \ and\ \bibinfo {author}
  {\bibfnamefont {B.~I.}\ \bibnamefont {Lundqvist}},\ }\bibfield  {title}
  {\enquote {\bibinfo {title} {{Van der Waals Density Functional for Layered
  Structures}},}\ }\href {\doibase 10.1103/PhysRevLett.91.126402} {\bibfield
  {journal} {\bibinfo  {journal} {Phys. Rev. Lett.}\ }\textbf {\bibinfo
  {volume} {91}},\ \bibinfo {pages} {126402} (\bibinfo {year}
  {2003})}\BibitemShut {NoStop}%
\bibitem [{\citenamefont {Gangopadhyay}\ \emph {et~al.}(1997)\citenamefont
  {Gangopadhyay}, \citenamefont {Jahanmir},\ and\ \citenamefont
  {Peterson}}]{gangopadhyay1997}%
  \BibitemOpen
  \bibfield  {author} {\bibinfo {author} {\bibfnamefont {A.}~\bibnamefont
  {Gangopadhyay}}, \bibinfo {author} {\bibfnamefont {S.}~\bibnamefont
  {Jahanmir}}, \ and\ \bibinfo {author} {\bibfnamefont {M.~B.}\ \bibnamefont
  {Peterson}},\ }\enquote {\bibinfo {title} {{S}elf-{L}ubricating {C}eramic
  {M}atrix {C}omposites},}\ \ (\bibinfo  {publisher} {Marcel Dekker},\ \bibinfo
  {year} {1997})\ Chap.~\bibinfo {chapter} {8}\BibitemShut {NoStop}%
\bibitem [{\citenamefont {Sansone}\ \emph {et~al.}(2016)\citenamefont
  {Sansone}, \citenamefont {Maschio}, \citenamefont {Usvyat}, \citenamefont
  {Schütz},\ and\ \citenamefont {Karttunen}}]{sansone2016}%
  \BibitemOpen
  \bibfield  {author} {\bibinfo {author} {\bibfnamefont {G.}~\bibnamefont
  {Sansone}}, \bibinfo {author} {\bibfnamefont {L.}~\bibnamefont {Maschio}},
  \bibinfo {author} {\bibfnamefont {D.}~\bibnamefont {Usvyat}}, \bibinfo
  {author} {\bibfnamefont {M.}~\bibnamefont {Schütz}}, \ and\ \bibinfo
  {author} {\bibfnamefont {A.}~\bibnamefont {Karttunen}},\ }\bibfield  {title}
  {\enquote {\bibinfo {title} {{Toward an Accurate Estimate of the Exfoliation
  Energy of Black Phosphorus: A Periodic Quantum Chemical Approach}},}\ }\href
  {\doibase 10.1021/acs.jpclett.5b02174} {\bibfield  {journal} {\bibinfo
  {journal} {The Journal of Physical Chemistry Letters}\ }\textbf {\bibinfo
  {volume} {7}},\ \bibinfo {pages} {131--136} (\bibinfo {year} {2016})},\
  \bibinfo {note} {pMID: 26651397}\BibitemShut {NoStop}%
\bibitem [{\citenamefont {Dong}\ \emph {et~al.}(2016)\citenamefont {Dong},
  \citenamefont {Zhang}, \citenamefont {Liu}, \citenamefont {Ye}, \citenamefont
  {Luo}, \citenamefont {Chen}, \citenamefont {Ma}, \citenamefont {Jie},
  \citenamefont {Chen}, \citenamefont {Wang},\ and\ \citenamefont
  {Zhang}}]{dong2016}%
  \BibitemOpen
  \bibfield  {author} {\bibinfo {author} {\bibfnamefont {S.}~\bibnamefont
  {Dong}}, \bibinfo {author} {\bibfnamefont {A.}~\bibnamefont {Zhang}},
  \bibinfo {author} {\bibfnamefont {J.}~\bibnamefont {Liu}, \bibfnamefont
  {K.~and~Ji}}, \bibinfo {author} {\bibfnamefont {Y.~G.}\ \bibnamefont {Ye}},
  \bibinfo {author} {\bibfnamefont {X.~G.}\ \bibnamefont {Luo}}, \bibinfo
  {author} {\bibfnamefont {X.~H.}\ \bibnamefont {Chen}}, \bibinfo {author}
  {\bibfnamefont {X.}~\bibnamefont {Ma}}, \bibinfo {author} {\bibfnamefont
  {Y.}~\bibnamefont {Jie}}, \bibinfo {author} {\bibfnamefont {C.}~\bibnamefont
  {Chen}}, \bibinfo {author} {\bibfnamefont {X.}~\bibnamefont {Wang}}, \ and\
  \bibinfo {author} {\bibfnamefont {Q.}~\bibnamefont {Zhang}},\ }\bibfield
  {title} {\enquote {\bibinfo {title} {{Ultralow-Frequency Collective
  Compression Mode and Strong Interlayer Coupling in Multilayer Black
  Phosphorus}},}\ }\href {\doibase 10.1103/PhysRevLett.116.087401} {\bibfield
  {journal} {\bibinfo  {journal} {Phys. Rev. Lett.}\ }\textbf {\bibinfo
  {volume} {116}},\ \bibinfo {pages} {087401} (\bibinfo {year}
  {2016})}\BibitemShut {NoStop}%
\bibitem [{\citenamefont {Liu}\ \emph {et~al.}(2012)\citenamefont {Liu},
  \citenamefont {Liu}, \citenamefont {Cheng}, \citenamefont {Li}, \citenamefont
  {Wang},\ and\ \citenamefont {Zheng}}]{liu2012}%
  \BibitemOpen
  \bibfield  {author} {\bibinfo {author} {\bibfnamefont {Z.}~\bibnamefont
  {Liu}}, \bibinfo {author} {\bibfnamefont {J.~Z.}\ \bibnamefont {Liu}},
  \bibinfo {author} {\bibfnamefont {Y.}~\bibnamefont {Cheng}}, \bibinfo
  {author} {\bibfnamefont {Z.}~\bibnamefont {Li}}, \bibinfo {author}
  {\bibfnamefont {L.}~\bibnamefont {Wang}}, \ and\ \bibinfo {author}
  {\bibfnamefont {Q.}~\bibnamefont {Zheng}},\ }\bibfield  {title} {\enquote
  {\bibinfo {title} {{Interlayer Binding Energy of Graphite: A Mesoscopic
  Determination from Deformation}},}\ }\href {\doibase
  10.1103/PhysRevB.85.205418} {\bibfield  {journal} {\bibinfo  {journal} {Phys.
  Rev. B}\ }\textbf {\bibinfo {volume} {85}},\ \bibinfo {pages} {205418}
  (\bibinfo {year} {2012})}\BibitemShut {NoStop}%
\bibitem [{\citenamefont {Gradt}\ and\ \citenamefont
  {Schneider}(2016)}]{gradt2016}%
  \BibitemOpen
  \bibfield  {author} {\bibinfo {author} {\bibfnamefont {T.}~\bibnamefont
  {Gradt}}\ and\ \bibinfo {author} {\bibfnamefont {T.}~\bibnamefont
  {Schneider}},\ }\bibfield  {title} {\enquote {\bibinfo {title}
  {{T}ribological {P}erformance of {M}o{S}2 {C}oatings in {V}arious
  {E}nvironments},}\ }\href {\doibase 10.3390/lubricants4030032} {\bibfield
  {journal} {\bibinfo  {journal} {Lubricants}\ }\textbf {\bibinfo {volume} {4}}
  (\bibinfo {year} {2016}),\ 10.3390/lubricants4030032}\BibitemShut {NoStop}%
\bibitem [{\citenamefont {Pope}(1976)}]{united1976friction}%
  \BibitemOpen
  \bibfield  {author} {\bibinfo {author} {\bibfnamefont {D.~P.}\ \bibnamefont
  {Pope}},\ }\href {\doibase https://doi.org/10.2172/7133423} {\emph {\bibinfo
  {title} {Friction and Wear Studies for Bonded Thin-film Lubricants}}},\
  \bibinfo {type} {Tech. Rep.}\ (\bibinfo  {institution} {{O}ffice of
  Scientific and Technical Information, U.S. Department of Energy},\ \bibinfo
  {year} {1976})\BibitemShut {NoStop}%
\bibitem [{\citenamefont {Wu}\ and\ \citenamefont {Schaden}(2014)}]{wu2014}%
  \BibitemOpen
  \bibfield  {author} {\bibinfo {author} {\bibfnamefont {H.~Y.}\ \bibnamefont
  {Wu}}\ and\ \bibinfo {author} {\bibfnamefont {M.}~\bibnamefont {Schaden}},\
  }\bibfield  {title} {\enquote {\bibinfo {title} {{Perturbative Roughness
  Corrections to Electromagnetic Casimir Energies}},}\ }\href {\doibase
  10.1103/PhysRevD.89.105003} {\bibfield  {journal} {\bibinfo  {journal} {Phys.
  Rev. D}\ }\textbf {\bibinfo {volume} {89}},\ \bibinfo {pages} {105003}
  (\bibinfo {year} {2014})}\BibitemShut {NoStop}%
\bibitem [{\citenamefont {Broer}\ \emph {et~al.}(2012)\citenamefont {Broer},
  \citenamefont {Palasantzas}, \citenamefont {Knoester},\ and\ \citenamefont
  {Svetovoy}}]{Broer2012}%
  \BibitemOpen
  \bibfield  {author} {\bibinfo {author} {\bibfnamefont {W.}~\bibnamefont
  {Broer}}, \bibinfo {author} {\bibfnamefont {G.}~\bibnamefont {Palasantzas}},
  \bibinfo {author} {\bibfnamefont {J.}~\bibnamefont {Knoester}}, \ and\
  \bibinfo {author} {\bibfnamefont {V.~B.}\ \bibnamefont {Svetovoy}},\
  }\bibfield  {title} {\enquote {\bibinfo {title} {{Roughness Correction to the
  Casimir Force at Short Separations: Contact Distance and Extreme Value
  Statistics}},}\ }\href {\doibase 10.1103/physrevb.85.155410} {\bibfield
  {journal} {\bibinfo  {journal} {Physical Review B}\ }\textbf {\bibinfo
  {volume} {85}} (\bibinfo {year} {2012}),\
  10.1103/physrevb.85.155410}\BibitemShut {NoStop}%
\bibitem [{\citenamefont {Maboudian}\ \emph {et~al.}(2000)\citenamefont
  {Maboudian}, \citenamefont {Ashurst},\ and\ \citenamefont
  {Carraro}}]{maboudian2000}%
  \BibitemOpen
  \bibfield  {author} {\bibinfo {author} {\bibfnamefont {R.}~\bibnamefont
  {Maboudian}}, \bibinfo {author} {\bibfnamefont {W.}~\bibnamefont {Ashurst}},
  \ and\ \bibinfo {author} {\bibfnamefont {C.}~\bibnamefont {Carraro}},\
  }\bibfield  {title} {\enquote {\bibinfo {title} {{Self-Assembled Monolayers
  as Anti-Stiction Coatings for MEMS: Characteristics and Recent
  Developments}},}\ }\href {\doibase 10.1016/S0924-4247(99)00337-4} {\bibfield
  {journal} {\bibinfo  {journal} {Sensors and Actuators A: Physical}\ }\textbf
  {\bibinfo {volume} {82}},\ \bibinfo {pages} {219--223} (\bibinfo {year}
  {2000})}\BibitemShut {NoStop}%
\end{thebibliography}%
\nocite{*} 

\end{document}